\documentclass[11pt]{article}
\usepackage{times}
\usepackage{graphicx}
\usepackage{amsfonts}
\usepackage{amssymb}
\usepackage{amsmath} 
\usepackage{amsthm}
\usepackage{natbib}
\usepackage{graphicx}

\newtheorem{proposition}{Proposition}
\newtheorem{definition}{Definition}
\newtheorem{lemma}{Lemma}
\newcommand{\OMIT}[1]{}

\newcommand {\ignore}[1]{}               

\newcommand {\br}[1]{\left(#1\right)}

\newcommand {\cbr}[1]{\left\{#1 \right\}}

\newcommand{\Mcal}{\mathcal{M}} 

\bibpunct{(}{)}{;}{a}{,}{,}

\begin{document}

\title{Graph kernels based on tree patterns for molecules}

\author{ Pierre Mah\'e \\
       Center for Computational Biology\\
       Ecole des Mines de Paris\\
       35, rue Saint Honor\'e, 77305 Fontainebleau, France \\
        \texttt{pierre.mahe@ensmp.fr }
       \and
        Jean-Philippe Vert \\
       Center for Computational Biology \\
       Ecole des Mines de Paris\\
       35, rue Saint Honor\'e, 77305 Fontainebleau, France \\
        \texttt{jean-philippe.vert@ensmp.fr} }


\maketitle

\begin{abstract}
Motivated by chemical applications, we revisit and extend a family of positive definite kernels for graphs based on the detection of common subtrees, initially proposed by \cite{Ramon2003Expressivity}. We propose new kernels with a parameter to control the complexity of the subtrees used as features to represent the graphs. This parameter allows to smoothly interpolate between classical graph kernels based on the count of common walks, on the one hand, and kernels that emphasize the detection of large common subtrees, on the other hand. We also propose two modular extensions to this formulation. The first extension increases the number of subtrees that define the feature space, and the second one removes noisy features from the graph representations. We validate experimentally these new kernels on binary classification tasks consisting in discriminating toxic and non-toxic molecules with support vector machines.
\end{abstract}


\section{Introduction}
There is an increasing need for algorithms to analyze and classify graph data, motivated in particular by various applications in chemoinformatics and bioinformatics. An prominent example in chemoinformatics, which motivates this work, is the generic problem of predicting various properties of small molecules, such as toxicological effects, given their {\em molecular graph}, that is, the graph representing the covalent bonds between atoms \citep{Leach2003introduction}. Classification of graphs is often associated with the problem of graph mining, which consists in detecting interesting patterns occurring in the graphs, and using them as features to build predictive models \citep{King1996Structure-activity,Inokuchi2003Complete,Helma2004Data,Deshpande2005Frequent}. 
As an alternative to this approach, kernel methods associated with graph kernels have recently emerged as a promising approach for classification of graph data. Kernel methods such as support vector machines (SVM) operate implicitly in a possibly high-dimensional Hilbert space of features, in the sense that no explicit computation of the image of the input data in the feature space is required. Instead, only the inner product between the images of any two input data points, called the \emph{kernel}, is required \citep{Scholkopf2002Learning,Shawe-Taylor2004Kernel}. Applying kernel methods to graph data therefore requires the definition of a kernel between graphs, thereafter simply referred to as \emph{graph kernel}. Choosing a graph kernel implicitly amounts to defining a set of features to represent the graphs and an inner product in the space of features.

Graph kernels were pioneered by \citet{Kashima2004Kernels} and \citet{Gartner2003graph}, who showed how to map graphs to an infinite-dimensional feature space indexed by linear subgraphs, and compute an inner product in that space. The resulting graph kernels compare two graphs through their common walks, weighted by a function of their lengths \citep{Gartner2003graph} or by their probability under a random walk model on the graphs \citep{Kashima2004Kernels}. While this representation might appear restrictive, these kernels led to promising empirical results, often comparing to state-of-the-art approaches in the fields of chemoinformatics \citep{Mahe2005Graph,Ralaivola2005Graph} and bioinformatics \citep{Borgwardt2005Protein,Karklin2005Classification}.

\begin{figure}[h]
        \begin{center}
        \includegraphics[width=7cm]{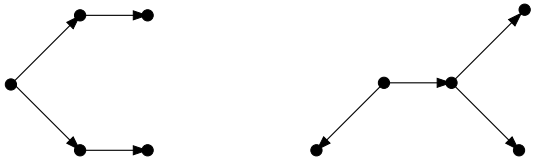}
        \caption{Two graphs having the same walk content, namely $\bullet : \times 5$ ; $\bullet \!\!\! \rightarrow \!\!\! \bullet : \times 4$ and  $\bullet \!\! \rightarrow \!\!\! \bullet \!\!\! \rightarrow \!\!\! \bullet : \times 2 $, and consequently mapped to the same point of the feature space corresponding a kernel based on the count of walks \citep{Gartner2003graph}. \label{fig:tree-identical} }
        \end{center}
\end{figure}

Nevertheless, \cite{Ramon2003Expressivity} highlighted the limited expressiveness of graph kernels based on linear features, showing in particular that many different graphs can be mapped to the same point in the corresponding feature space. Figure \ref{fig:tree-identical} illustrates this issue on a simple example. On the other hand, they also showed that computing a perfect graph kernel, that is, a kernel mapping non-isomorphic graphs to distinct points in the feature space, is NP-hard. This suggests that the expressiveness of graph kernels must be traded for their computational complexity. As a first step towards a refinement of the feature space used in walk-based graph kernels, \cite{Ramon2003Expressivity} introduced a kernel function comparing graphs on the basis of their common subtrees. This representation looks promising in particular in chemoinformatics, because physicochemical properties of atoms are known to be related to their topological environment that could be well captured by subtrees. However, the relationship between the new subtree-based kernel and previous walk-based kernels was not analyzed in details, and the relevance of the new kernel was not tested empirically.

Our motivation in this paper is to study in detail, both theoretically and empirically, the relevance of subtree features for graph kernels, and in particular to assess the benefits they bring compared to walk-based graph kernels. For that purpose we first revisit the formulation introduced by \citet{Ramon2003Expressivity} and propose two new kernels with an explicit description of their feature spaces and corresponding inner products. We introduce a parameter in the formulations that allows to gradually increase the complexity of the subtrees used as features to represent the graphs, the notion of complexity depending on the formulation. By decreasing the parameter we recover classical walk-based kernels, and by increasing it, we can empirically observe in detail the effect of increasing the number and the complexity of the tree features used to represent the graphs. Both formulations can be efficiently computed by dynamic programming, in the spirit of the kernel proposed by \citet{Ramon2003Expressivity}. When the size of allowed subtrees is increased, however, we observe that the practical use of this kernel is limited by the explosion in the number of subtrees occurring in the graphs. In a second step, we therefore introduce two extensions to the initial formulation of the kernels that allow, on the one hand, to extend and generalize their associated feature space, and on the other hand, to remove noisy features that correspond to unwanted subtrees. The different kernels are compared experimentally on two binary classification tasks consisting in discriminating toxic from non-toxic molecules with a SVM.

Although our main motivations are in chemical applications, we adopt the general framework of graph kernels in this paper, because the kernels introduced may find different applications in domains where data have a natural graph structure, such as bioinformatics, natural language processing or image processing. We assume that the reader is familiar with kernel functions and SVMs, and refer him to \cite{Scholkopf2002Learning,Shawe-Taylor2004Kernel} and references therein for a background on the subject. The remaining of the paper is organized as follows. Notations and definitions related to graphs and trees are introduced in Section \ref{sec:prelim}, followed in Section \ref{sec:definition} by the definition of a general class of kernels based on the detection of common subtrees. The next section (Section \ref{sec:gartner}) revisits the framework introduced in \cite{Ramon2003Expressivity}, from which two particular graph kernels are derived and further extended in Section \ref{sec:extensions}.
The kernels are validated experimentally in Section \ref{sec:experiments}, and we give concluding remarks in Section \ref{sec:discussion}.

\section{Notations and Definitions}\label{sec:prelim}

In this section we introduce notations and general definitions related to graphs and trees.

\subsection{Labeled Directed Graphs}
A {\em labeled graph} $G = (\mathcal{V}_G,\mathcal{E}_G)$ is defined by a finite set of {\em vertices} $\mathcal{V}_G$,
a set of {\em edges} $\mathcal{E}_G \subset \mathcal{V}_G \times \mathcal{V}_G$, and a labeling function $l: \mathcal{V}_G \, \cup \, \mathcal{E}_G \rightarrow \mathcal{A}$ which assigns a {\em label} $l(x)$ taken from an alphabet $\mathcal{A}$ to any vertex or edge $x$.
We let $|\mathcal{V}_G|$ be the number of vertices of $G$, $|\mathcal{E}_G|$ be its number of edges, and we assume below that a set of labels $\mathcal{A}$ common to all graphs has been fixed.
In {\em directed} graphs, edges are oriented and to each vertex $u \in \mathcal{V}_G$ corresponds a set of {\em incoming neighbors} $\delta^-(u)= \{ v \in \mathcal{V}_G: (v,u) \in \mathcal{E}_G \}$ and {\em outgoing neighbors} $\delta^+(u) = \{ v \in \mathcal{V}_G: (u,v) \in \mathcal{E}_G \}$.
We let $d^-(u) = |\delta^-(u)|$ be the {\em in-degree} of the vertex $u$, and $d^+(u) = |\delta^+(u)|$ be its {\em out-degree}.
A {\em walk} of length $n$ in the graph $G = (\mathcal{V}_G,\mathcal{E}_G)$ is a succession of $n+1$ vertices $(v_0,\dots,v_n) \in \mathcal{V}_G^{n+1}$, such that $(v_i,v_{i+1}) \in \mathcal{E}_G$ for $i=0,\ldots,n-1$.
A {\em path} is a walk $(v_0,\dots,v_n)$ with the additional condition that $i \neq j \iff v_i \neq v_j$.
Finally, a graph is said to be {\em connected} if there is a walk between any pair of vertices when the orientation of edges is dropped.

\begin{figure}[h]
        \begin{center}
        \includegraphics[width=7cm]{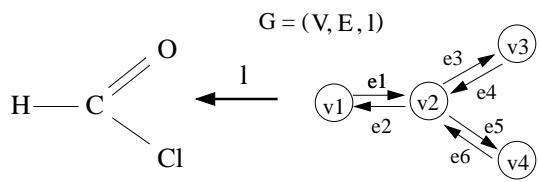}
        \caption{A chemical compound seen as a labeled graph \label{fig:labeled} }
        \end{center}
\end{figure}

For applications in chemistry considered below, we associate a labeled directed graph $G = (\mathcal{V}_G,\mathcal{E}_G)$ to the planar structure of a molecule.
To do so, we let the set of vertices $\mathcal{V}_G$ correspond to the set of atoms of the molecule, the set of edges $\mathcal{E}_G$ to its covalent bonds, and label these graph elements according to an alphabet $\mathcal{A}$ consisting of the different types of atoms and bonds. 
Note that since graphs are directed, a pair of edges of opposite direction is introduced for each covalent bond of the molecule. Figure \ref{fig:labeled} shows a chemical compound seen as a labeled directed graph.

\subsection{Trees}

A {\em tree} $t$ is a directed connected acyclic graph in which all vertices have in-degree one, except one that has in-degree zero.
The node with in-degree zero is known as the {\em root} $r(t)$ of the tree. Nodes with out-degree zero are known as {\em leaf} nodes, others are called {\em internal} nodes. Trees are naturally oriented, edges being directed from the root to the leaves. The outgoing neighbors of an internal node are known as its {\em children}, and the unique incoming neighbor of a node (apart from the root) is known as its {\em parent}. If two nodes have the same parent, their are said to be {\em siblings}. The {\em size} $|t|$ of the tree $t$ is its number of nodes: $|t| = |\mathcal{V}_t|$. The {\em depth} of a node corresponds to the number of edges connecting it to the root plus one\footnote{Note that the depth of the root node is one.}, and the depth of the tree is the maximum depth of its nodes. Finally, we introduce a couple of definitions that will be useful in the following.

\begin{definition}[Balanced tree]\label{def:balanced-tree}
A {\em perfectly depth-balanced tree} of order $h$ is a tree where the depth of each leaf node is $h$. Perfectly depth-balanced trees are also called {\em balanced trees} below.
\end{definition}

\begin{figure}[h]
        \begin{center}
        \includegraphics[width=7cm]{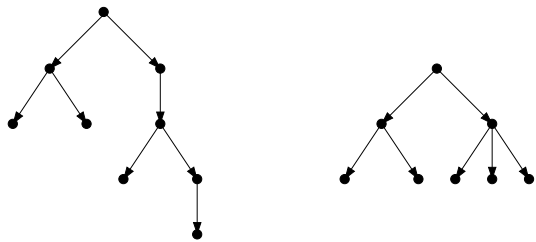}
        \caption{Left: a tree $t_1$ of depth 5 with $|t_1| = 9$ and $\text{branch}(t_1) = 3$. 
        Right: a balanced tree $t_2$ of order 3 with $|t_2| = 8$ and $\text{branch}(t_2) = 4$.
        Top nodes are root nodes, bottom nodes are leaf nodes. \label{fig:tree} }
        \end{center}
\end{figure}

\begin{definition}[Branching cardinality]\label{def:branching}

We define the {\em branching cardinality} of the tree $t$, noted $\text{branch}(t)$, as its number of leaf nodes minus one. More formally, for the tree $t = (\mathcal{V}_t,\mathcal{E}_t)$ with $\mathcal{V}_t = (v_1,\dots,v_{|t|})$, $\text{branch}(t)$ is given by;
\begin{equation*}
\text{branch}(t) = \sum_{i=1}^{|t|} {\bf 1}(d^+(v_i)=0) - 1,
\end{equation*}
where ${\bf 1}(.)$ is a binary function equal to one if its argument is true, and zero otherwise.
\end{definition}
This terminology stems from the observation that this quantity also corresponds to the sum, over the non-leaf nodes of the tree, of their numbers of children minus one. It therefore measures how many extra branchings there are compared to a linear tree, which has branching cardinality 0.  These definitions are illustrated in Figure \ref{fig:tree}.\\

The remaining of the paper introduces kernel functions between labeled directed graphs based on the detection in the graphs of patterns corresponding to labeled trees. To lighten notations, we simply refer below to labeled directed graphs and labeled trees as graphs and trees.

\section{The Tree-Pattern Graph Kernel}\label{sec:definition}
This section introduces a general class of graph kernel based on the detection, in the graphs, of patterns corresponding to particular tree structures. We start by defining precisely this notion of tree-pattern.
\begin{definition}[Tree-pattern]\label{def:tree-pattern}
Let a graph $G= (\mathcal{V}_G,\mathcal{E}_G)$ and a tree $t= (\mathcal{V}_t,\mathcal{E}_t)$, with $\mathcal{V}_t = (n_1,\dots,n_{|t|})$. A $|t|$-uple of vertices $(v_1,\dots, v_{|t|}) \in \mathcal{V}_G^{|t|}$ is a {\em tree-pattern} of $G$ with respect to $t$, which we denote by $(v_1,\dots, v_{|t|}) = \text{pattern}(t)$, if and only if the following holds:
\begin{equation*}
\begin{cases}
 \forall i \in [1,|t|], &l(v_{i}) = l(n_i)\,, \\
 \forall (n_i,n_j) \in \mathcal{E}_t, & (v_{i},v_{j}) \in \mathcal{E}_G \wedge l\big((v_{i},v_{j})\big) = l\big( (n_i,n_j)\big)\,, \\
\forall (n_{i}, n_{j}), (n_{i},n_{k}) \in \mathcal{E}_t, &  j \neq k \iff v_{j} \neq v_{k} \,.
\end{cases}
\end{equation*}
\end{definition}
In other words a tree-pattern is a combination of graph vertices that can be arranged in a particular tree structure, according to the labels and the connectivity properties of the graph.
Note from this definition that vertices of the graph are allowed to appear several times in a tree-pattern, under the condition that siblings nodes of the corresponding tree are associated to distinct vertices of the graphs.
We now introduce a functional to count occurrences of these patterns.
\begin{definition}[Tree-pattern counting function]\label{def:tree-counting}
A \emph{tree-pattern counting function} returning the number of times a tree-pattern occurs in a graph is defined for the tree $t$ and the graph 
$G= (\mathcal{V}_G,\mathcal{E}_G)$, $\mathcal{V}_G = (v_1,\dots,v_{|\mathcal{V}_G|})$, as
\begin{equation*}
\psi_t(G) = \big\vert \big\{  (\alpha_1,\dots,\alpha_{|t|}) \in [1,|\mathcal{V}_G|]^{|t|} : 
(v_{\alpha_1},\dots, v_{\alpha_{|t|}}) = \text{pattern}(t)  \big\}  \big\vert.
\end{equation*}
A restriction of $\psi_t$ to patterns rooted in a specified vertex $v$ is given by
\begin{equation*}
\psi_t^{(v)}(G) = \big\vert \big\{ (\alpha_1,\dots,\alpha_{|t|})  \in [1,|\mathcal{V}_G|]^{|t|} : \; (v_{\alpha_1},\dots, v_{\alpha_{|t|}}) = \text{pattern}(t) \wedge v_{\alpha_1} = v \big\} \big\vert.
\end{equation*}
\end{definition}
With this new definition at hand we can define a general graph kernel based on the detection of common tree-patterns in the graphs.
\begin{definition}[Tree-pattern graph kernel]\label{def:kernel}
{\em The tree-pattern graph kernel} $K$ is given for the graphs $G_1$ and $G_2$ by 
\begin{equation*}
K(G_1,G_2) = \sum_{t \in \mathcal{T}} w(t) \psi_t(G_1) \psi_t(G_2),
\end{equation*}
where $\mathcal{T}$ is a set of trees, $w : \mathcal{T} \rightarrow \mathbb{R}$ is a tree weighting functional and $\psi_t$ is the tree-pattern counting function of Definition \ref{def:tree-counting}.
\end{definition}

\begin{figure}[h]
        \begin{center}
        \includegraphics[width=10cm]{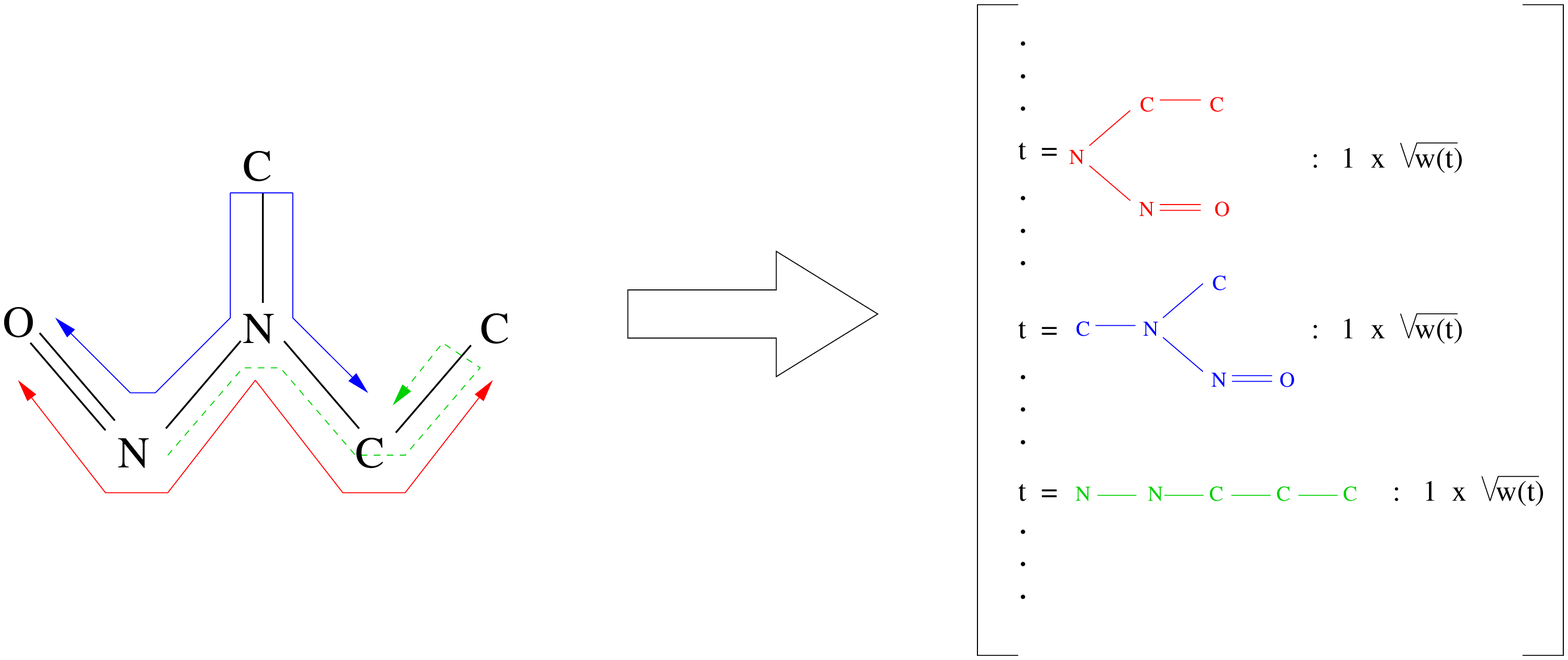}
        \caption{A molecular compound $G$ (left) and its feature space representation $\phi(G)$ (right). Note that the red and green trees are balanced. Note moreover that the green tree consists of a set of linearly connected atoms, which is known as {\em molecular fragment} in chemoinformatics. Note finally that the same $C$ atom appears in the 3rd and 5th positions in the tree-pattern corresponding to the green tree.\label{fig:patterns} }
        \end{center}
\end{figure}

The kernel of Definition \ref{def:kernel} is obviously positive definite since it can be written as a standard dot-product $K(G_1,G_2) = \langle \phi(G_1),\phi(G_2) \rangle$, where $\phi(G)$ is the mapping that maps any graph $G$ to the feature space indexed by the trees of the set $\mathcal{T}$ as $\phi(G) = \big( \sqrt{w(t)} \psi_t(G) \big)_{t \in \mathcal{T}}$.
Figure \ref{fig:patterns} illustrates this mapping.

\section{Examples of tree-pattern graph kernels}\label{sec:gartner}
In a recent work, \citet{Ramon2003Expressivity} proposed a particular tree-pattern graph kernel fitting the general Definition \ref{def:kernel}. In this section, we propose two different kernels with explicit feature spaces and inner products, discuss their practical computation, and highlight their differences with the kernel of \citet{Ramon2003Expressivity}.

        \subsection{Kernels Definition}\label{sec:gartner-def}

According to Definition \ref{def:kernel}, two key elements enter in the definition of a tree-pattern graph kernel. Firstly, the set of trees  $\mathcal{T}$ indexing the feature space the graphs are mapped to must be chosen. The kernels we consider in this section are based on the same feature space: the space indexed by the set of balanced trees of order $h$ introduced in Definition \ref{def:balanced-tree}, labeled according to the graphs labeling alphabet $\mathcal{A}$. We will refer to this set as $\mathcal{B}_h$ in the following. Second, the tree weighting function $w$ must be defined. A natural way to define such a functional is to take into account the structure of the trees, and accordingly, we propose to relate the weight of a tree to its size or its branching cardinality. In particular we propose to consider the following kernels:
\begin{definition}[Size-based balanced tree-pattern kernel]\label{def:kernel-size}
For the pair of graphs $G_1$ and $G_2$, the {\em size-based balanced tree-pattern kernel of order $h$} is defined as
\begin{equation}\label{eq:kernel-size}
K_{\text{Size}}^h(G_1,G_2) = \sum_{t \in \mathcal{B}_h} \lambda^{|t|-h} \psi_t(G_1) \psi_t(G_2).
\end{equation}
\end{definition}

\begin{figure}[h]
        \begin{center}
        \includegraphics[width=13cm]{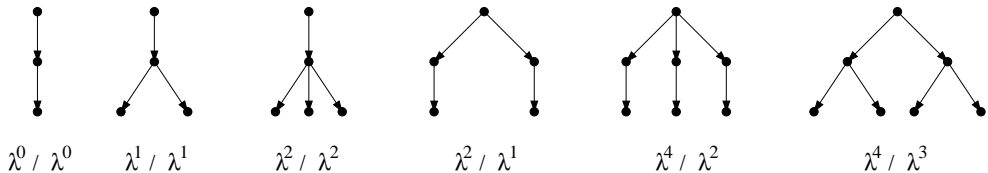}
        \caption{A set of balanced trees of order 3, together with their size-based (left) and branching-based (right) $\lambda$ weighting.\label{fig:weights} }
        \end{center}
\end{figure}

\begin{definition}[Branching-based balanced tree-pattern  kernel]\label{def:kernel-branch}
For the pair of graphs $G_1$ and $G_2$, the {\em branching-based balanced tree-pattern kernel of order $h$} is defined as
\begin{equation}\label{eq:kernel-branch}
K_{\text{Branch}}^h(G_1,G_2) = \sum_{t \in \mathcal{B}_h} \lambda^{\text{branch}(t)} \psi_t(G_1) \psi_t(G_2).
\end{equation}
\end{definition}
Note that the depth of a tree is a lower bound on its size, attained for a tree consisting of a linear chain of vertices. For such a tree, at depth $h$, we have $|t|-h = \text{branch}(t)=0$, and we see that the corresponding tree-patterns are given a unit weight in the kernels of Definitions \ref{def:kernel-size} and \ref{def:kernel-branch}.
The complexity of a tree naturally increases with its size and  branching cardinality, and the $\lambda$ parameter entering the kernel Definitions \ref{def:kernel-size} and \ref{def:kernel-branch} has the effect of favoring tree-patterns depending on their degree of complexity. A value of $\lambda$ greater than one favors the influence of tree-patterns of increasing complexity over the trivial linear tree-patterns, while they are penalized by a value of $\lambda$ smaller than one. We can note, however, that while the size of a tree increases with its branching cardinality, the converse is not true.
For any tree $t$ of depth $h$, we therefore always have $|t|-h \geq \text{branch}(t)$, and the tree weighting is more important in the size-based than in the branching-based kernel. In the case of balanced trees, this difference is particularly marked when the nodes with large out-degree are close to the root node. This is due to the fact that every leaf must be at depth $h$, and while the size of the tree necessarily increases by at least $h-1$ along each path starting from the root, the branching cardinality does not\footnote{At the extreme, we have $|t| = 1 + (h-1) \times d^+(r(t))$ Vs $\text{branch}(t) = d^+(r(t)) -1$.}. The main difference in the feature space representations of the graphs is therefore induced by this particular type of tree-patterns, that can be interpreted as collections of regular subtree patterns merged in the root node. This suggests for instance that, for $\lambda < 1$, the branching-based formulation of the kernel may to some extent tolerate large, yet regular patterns, that would be strongly penalized in the size-based formulation. Figure \ref{fig:weights} illustrates these tree weightings based on the size and branching cardinality.\\

When $\lambda$ tends to zero, the complexity of the patterns is so penalized that only tree-patterns consisting of linear chains of graph vertices have non-vanishing weights, and the kernels of Definitions \ref{def:kernel-size} and \ref{def:kernel-branch} boil down to a kernel based on the detection of common walks \citep{Gartner2003graph}. More formally, if we define the set of walks of length $n$ of the graph $G$ as 
$$\mathcal{W}_n(G) = \{ (v_0,\dots,v_n) \in \mathcal{V}_G^{n+1}: (v_i,v_{i+1}) \in \mathcal{E}_G, \; 0 \leq i \leq n-1 \}, $$
and define for the graphs $G_1$ and $G_2$ the following walk-count kernel:
\begin{equation}\label{eq:kernel-walk}
K_{\text{Walk}}^{n}(G_1,G_2) = \sum_{\substack{w_1 \in \\ \mathcal{W}_{n}(G_1)}} \sum_{\substack{w_2 \in \\ \mathcal{W}_{n}(G_2)}} {\bf 1}(l(w_1)=l(w_2)),
\end{equation}
where ${\bf 1}(l(w_1)=l(w_2))$ is one if all pairs of corresponding edges and vertices are identically labeled in the walks $w_1$ and $w_2$, and zero otherwise, one easily gets that:
\begin{equation*}
\lim_{\lambda \rightarrow 0} K_{\text{Size}}^h(G_1,G_2) = \lim_{\lambda \rightarrow 0} K_{\text{Branch}}^h(G_1,G_2) = K_{\text{Walk}}^{h-1}(G_1,G_2).
\end{equation*}
Increasing the value of $\lambda$ relaxes the penalization on complex subtree features, and can therefore be interpreted as introducing tree-patterns of increasing complexity in the walk-based kernel of Equation \ref{eq:kernel-walk}.\\

It should be noted finally that the parameters $h$ and $\lambda$ are directly related to the nature of the features representing the graphs and to their relative importance. Optimal values of the parameters are therefore likely to be dependent on the problem and data considered, and can hardly be chosen a priori. As an example, because of the variety of chemical compounds, the graphs considered in a chemical application can have a great structural diversity. This suggests that these parameters should be estimated from the data using, for example, cross-validation techniques.

        \subsection{Kernels Computation}\label{sec:compute}
We now propose two factorization schemes to compute the kernels of Definitions \ref{def:kernel-size} and \ref{def:kernel-branch}. These factorizations are inspired by the dynamic programming (DP) algorithm proposed by \cite{Ramon2003Expressivity} to compute a slightly different graph kernel, discussed in the next subsection. The factorization relies on the following definition:

\begin{definition}[Neighborhood matching set]\label{def:matching}
The \emph{neighborhood matching set} $\mathcal{M}(u,v)$ of two graph vertices $u$ and $v$ is defined as
\begin{align*}
\mathcal{M}(u,v) = \big\{ R \subseteq \delta^+(u) \times \delta^+(v) \;|\; & \big( \forall (a,b), (c,d) \in R : a \neq c \wedge b \neq d \big) \\
& \wedge \big( \forall (a,b) \in R : l(a) = l(b) \wedge l((u,a)) = l((v,b)) \big) \big\}.
\end{align*}
\end{definition}
Each $ R \in \mathcal{M}(u,v)$ consists of one or several pair(s) of neighbors of $u$ and $v$ that are identically labeled and connected to $u$ and $v$ by edges of the same label.
It follows from Definition \ref{def:balanced-tree} that such an element $R$ corresponds to a pair of balanced tree-patterns of order 2 rooted in $u$ and $v$, found in the graph(s) $u$ and $v$ belong to. Moreover, provided $u$ and $v$ have the same label, these patterns correspond to the same balanced tree. We can state the following propositions, whose proofs are post-poned in Appendix \ref{app:proof1}:

\begin{proposition}[Size-based kernel computation]\label{prop:comput-size}
The order $h$ size-based tree-pattern kernel $K_{\text{Size}}^h$ of Definition \ref{def:kernel-size} between two graphs $G_1$ and $G_2$ can be computed as:
\begin{equation}\label{eq:comput-size}
K_{\text{Size}}^h(G_1,G_2) = \frac{1}{\lambda^h} \sum_{u \in \mathcal{V}_{G_1}} \sum_{v \in \mathcal{V}_{G_2}} k_{h}(u,v),
\end{equation}
where $k_n, n=1,\ldots,h$ is defined recursively by
\begin{equation*}
\begin{cases}
k_{1}(u,v) = \lambda {\bf 1}(l(u) = l(v)) \,,\\
\displaystyle k_{n}(u,v) = \lambda {\bf 1}(l(u) = l(v)) \sum_{R \in \mathcal{M}(u,v)} \prod_{(u',v') \in R} k_{n-1}(u',v'), \quad n=2,\dots,h. \\
\end{cases}
\end{equation*}

\end{proposition}

\begin{proposition}[Branching-based kernel computation]\label{prop:comput-branch}
The order $h$ branching-based tree-pattern kernel $K_{\text{Branch}}^h$ of Definition \ref{def:kernel-branch} between two graphs $G_1$ and $G_2$ can be computed as:
\begin{equation}\label{eq:comput-branch}
K_{\text{Branch}}^h(G_1,G_2) = \sum_{u \in \mathcal{V}_{G_1}} \sum_{v \in \mathcal{V}_{G_2}} k_{h}(u,v),
\end{equation}
where $k_n, n=1,\ldots,h$ is defined recursively by
\begin{equation*}
\begin{cases}
k_{1}(u,v) = {\bf 1}(l(u) = l(v))\,, \\
\displaystyle k_{n}(u,v) = {\bf 1}(l(u) = l(v)) \sum_{R \in \mathcal{M}(u,v)} \frac{1}{\lambda} \prod_{(u',v') \in R} \lambda k_{n-1}(u',v'), \quad n=2,\dots,h.
\end{cases}
\end{equation*}
\end{proposition}
Not surprisingly, Propositions \ref{prop:comput-size} and \ref{prop:comput-branch} show that the kernels $K_{\text{Size}}^h$ and $K_{\text{Branch}}^h$ of Definitions \ref{def:kernel-size} and \ref{def:kernel-branch} have the same complexity.
More precisely, for the pair of graphs $G_1$ and $G_2$, it follows from (\ref{eq:comput-size}) and (\ref{eq:comput-branch}) that this complexity is equal to the product of the sizes of $G_1$ and $G_2$, times the complexity of evaluating the functional $k_h$.
In both cases, for the pair of graph vertices $u$ and $v$, evaluating $k_h(u,v)$ amounts to summing, over all possible matching of neighbors $R \in \mathcal{M}(u,v)$, a quantity expressed as a product of $|R|$ functionals $k_{h-1}$.
The size of $\mathcal{M}(u,v)$, $|\mathcal{M}(u,v)|$, is maximal if all the neighbors of $u$ and $v$, as well as the edges that connect them to $u$ and $v$, are identically labeled.
In that case we have
$$
|\mathcal{M}(u,v)| = \sum_{k=1}^{\min( d^+(u),  d^+(v) )} A^k_{ d^+(u)} A^k_{ d^+(v)},
$$
where $k$ ranges over the cardinality $|R|$ of the set of matching neighbors.
If we let $d$ be an upper bond on the out-degree of the vertices of the graphs considered, it follows that 
$
|\mathcal{M}(u,v)| \leq \sum_{k=1}^{d} (A^k_d)^2
$
and we can derive the following worst case complexity
$$
\mathcal{O}(K_{\text{Size}}^h(G_1,G_2)) = \mathcal{O}(K_{\text{Branch}}^h(G_1,G_2))  =  |\mathcal{V}_{G_1}| \times |\mathcal{V}_{G_2}| \times ( \sum_{k=1}^d k(A_d^k)^2 )^{h-1}.
$$
In the case of chemical compounds, we have $d=4$. The factor $ \sum_{k=1}^d k(A_d^k)^2$ equals 4336, and the complexity looks prohibitive. However this is only a worst-case complexity which is strongly reduced in practice because (i) the out-degree of the vertices is often smaller than 4\footnote{For example, in the two datasets considered in our experiments in section \ref{sec:experiments}, the average out-degree of the vertices is nearly 2 (2.14 for the first dataset, and 2.06 for the second one).}, and (ii) the size of $\mathcal{M}(u,v)$ is reduced by the fact that vertices and edges can have distinct labels.\\

\subsection{Relation to previous work}
At this point, it is worth reminding the kernel formulation introduced by \cite{Ramon2003Expressivity} in order to highlight the differences with the kernels proposed in Definitions \ref{def:kernel-size} and \ref{def:kernel-branch}. In the context of graphs with labeled vertices and edges\footnote{The original formulation considered graphs with labeled vertices only, and the definition of the neighborhood matching set is refined in this paper in order to handle labeled edges.}, at order $h$, the kernel introduced in \cite{Ramon2003Expressivity}, that we denote by $K_{\text{Ramon}}^h$, is formulated as follows:

\begin{equation*}
K_{\text{Ramon}}^h(G_1,G_2) = \sum_{u \in \mathcal{V}_{G_1}} \sum_{v \in \mathcal{V}_{G_2}} k_{h}(u,v),
\end{equation*}
where $k_n$ is defined by
\begin{equation*}
\begin{cases}
k_{1}(u,v) = {\bf 1}(l(u) = l(v)) \\
\displaystyle k_{n}(u,v) = {\bf 1}(l(u) = l(v)) \; \lambda_u \lambda_v  \sum_{R \in \mathcal{M}(u,v)} \prod_{(u',v') \in R} k_{n-1}(u',v'), \quad n=2,\dots,h. 
\end{cases}
\end{equation*}
It is clear that this kernel and the kernels of Definitions \ref{def:kernel-size} and \ref{def:kernel-branch} have the same feature space. The main difference lies in the fact that in this formulation, a parameter $\lambda_v$ is introduced for each vertex $v$ of each graph. It can be checked that under this parametrization, each tree-pattern is weighted by the product of the parameters $\lambda_v$ associated to its internal nodes. In the special case where these parameters are taken equal to a single parameter $\lambda$, each pattern is therefore weighted by $\lambda$ raised to the power of its number of internal nodes. While this bears some similarity with the size-based weighting proposed in the kernel of Definition \ref{def:kernel-size}, we note for instance that the three leftmost trees of Figure \ref{fig:weights} are identically weighted, namely by a factor $\lambda^2$. Moreover, the convergence to the walk-based kernel of Equation \ref{eq:kernel-walk} observed when $\lambda$ tends to zero for the kernels of Definition  \ref{def:kernel-size} and \ref{def:kernel-branch} does not hold with this formulation.

\section{Extensions}\label{sec:extensions}

The kernels introduced in the previous section arise directly from the adaptation of the algorithm proposed in \cite{Ramon2003Expressivity}.
In this section we introduce two extensions to this initial formulation.
First, we extend the branching-based kernel of Definition \ref{def:kernel-branch} to a feature space indexed by a larger, and more general, set of trees.
Second, we propose to eliminate a set of noisy tree-patterns from the feature space.
        
        \subsection{Considering all trees}\label{sec:untilN}
The DP algorithms of Section \ref{sec:compute} recursively extend the tree-patterns under construction until they reach a specified depth.
Because they are based on the notion of neighborhood matching sets introduced in Definition \ref{def:matching}, these algorithms add at least one child to every leaf node of the patterns under extension at each step of the recursive process.
When they reach the specified depth, the patterns are therefore balanced, and the choice of the feature space associated to the kernels of Definitions \ref{def:kernel-size} and \ref{def:kernel-branch} was actually dictated by their computation.

Rather than focusing on features of a particular size, standard representations of molecules involve structural features of different sizes. A prominent example is that of molecular fingerprints \citep{Ralaivola2005Graph} that typically represent a molecule by its exhaustive list of fragments of length up to 8, where a fragment is defined as a linear succession of connected atoms (see Figure \ref{fig:patterns}). In this section, we note that a slight modification of the DP algorithm of Proposition \ref{prop:comput-branch} generalizes the kernel of Definition \ref{def:kernel-branch} to a feature space indexed by the set of general trees up to a given depth, instead of the set of balanced-trees of the corresponding order.
More precisely, if we let $\mathcal{T}_h$ be the set of trees of depth up to $h$, and if we define the {\em until-N extension} of the branching-based kernel of Definition \ref{def:kernel-branch} as
\begin{equation}\label{eq:kernel-untilN}
K_{\text{Branch}}^{\text{until-}h}(G_1,G_2) = \sum_{t \in \mathcal{T}_h} \lambda^{\text{branch}(t)} \psi_t(G_1) \psi_t(G_2),
\end{equation}
we can state the following proposition, whose proof is postponed in Appendix \ref{app:proof-untilN}.

\begin{figure}[h]
\begin{center}
    \includegraphics[width=12cm]{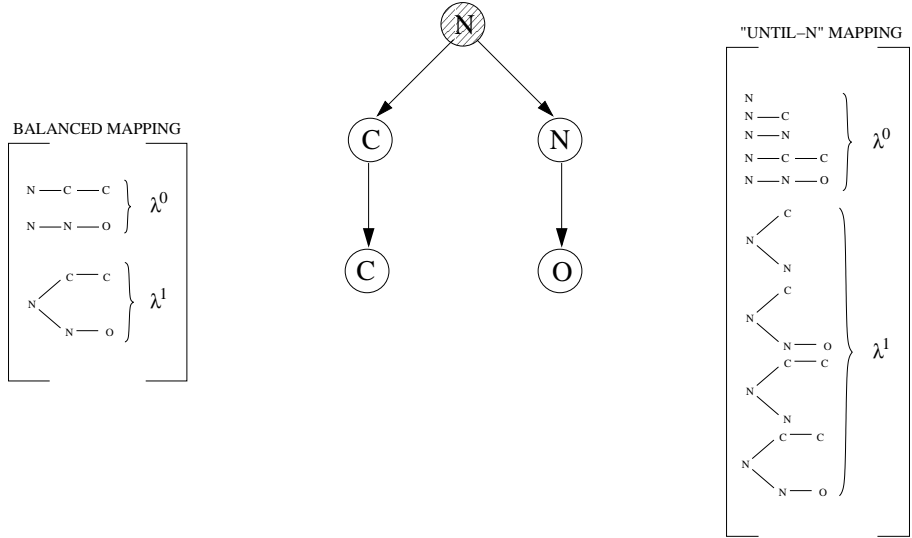}
    \caption{A graph $G$, and the set of  balanced trees of order 3 (left) and general trees of depth up to 3 (right) for which a tree-pattern rooted in the dashed vertex is found in $G$, together with their kernel weighting $\lambda^{\text{branch}(t)}$.\label{fig:mappings} }
\end{center}
\end{figure}

\begin{proposition}[Until-N  kernel computation]\label{prop:untilN}
The {\em until-N extension} $K_{\text{Branch}}^{\text{until-}h}$ of the branching-based kernel of order $h$ of Definition \ref{def:kernel-branch} is given for the graphs $G_1$ and $G_2$ by
$$
K_{\text{Branch}}^{\text{until-}h}(G_1,G_2) = \sum_{u \in \mathcal{V}_{G_1}} \sum_{v \in \mathcal{V}_{G_2}} k_{h}(u,v),
$$
where $k_n, n=1,\ldots,h$ is defined recursively by
\begin{equation*}
\begin{cases}
k_{1}(u,v) = {\bf 1}(l(u) = l(v)) \,, \\
\displaystyle k_{n}(u,v) = {\bf 1}(l(u) = l(v)) \left( 1 + \sum_{R \in \mathcal{M}(u,v)} \frac{1}{\lambda} \prod_{(u',v') \in R} \lambda k_{n-1}(u',v') \right), \quad n=2,\dots,h.
\end{cases}
\end{equation*}
\end{proposition}
The computation given in Proposition \ref{prop:untilN} follows that of Proposition \ref{prop:comput-branch}, and this until-N extension comes at no extra cost. The feature space corresponding to this extended kernel has nevertheless a much larger dimensionality than that of the original branching-based kernel. Actually, because the set of trees $\mathcal{T}_h$ includes the set of balanced trees $\mathcal{B}_h$ as a special case, the feature space associated to the branching-based kernel is a sub-space of the feature space associated to its until-N extension. Figure \ref{fig:mappings} illustrate the different mappings.
The behavior of this kernel with respect to $\lambda$ follows that of the original branching-based kernel. In particular, when $\lambda$ tends to zero, the set of tree-patterns with non-vanishing weights reduces to linear chain of vertices and the kernel boils down to a kernel based on the detection of common walks of length up to $h-1$. More formally, one can easily check that, in this case:
$$
\lim_{\lambda \rightarrow 0} K_{\text{Branch}}^{\text{until-}h}(G_1,G_2) = \sum_{n = 0}^{h-1} K_{\text{Walk}}^{n}(G_1,G_2),
$$
where $K_{\text{Walk}}^{n}$ is the kernel based on the detection of common walks of length $n$, defined in Section \ref{sec:gartner-def}, Equation \ref{eq:kernel-walk}.

Finally, we note that this extension is not directly applicable to the size-based kernel of Definition \ref{def:kernel-size} because of a slight difference in the computations of Propositions \ref{prop:comput-size} and \ref{prop:comput-branch}.
Indeed, note from Proposition \ref{prop:comput-size} that in order to get the $\lambda^{|t|-h}$ weighting of the tree $t$ proposed in Definition \ref{def:kernel-size}, the size-based kernel is initially computed from patterns weighted by their sizes, and is subsequently normalized by a factor $\lambda^{-h}$.
As a result, while the above extension would still have the effect of extending the feature space to the space indexed by trees of $\mathcal{T}_h$, this $\lambda^{-h}$ normalization would affect every tree-pattern regardless of their size, and the pattern weighting proposed in Definition \ref{def:kernel-size} would be lost.

        \subsection{Removing tottering tree-patterns}\label{sec:nototters}
\begin{figure}[h]
        \begin{center}
        \includegraphics[width=8cm]{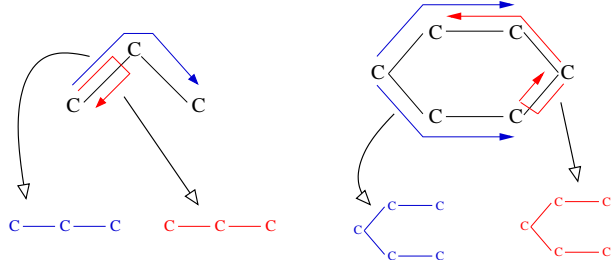}
        \caption{Left: tottering (red) and no-tottering (blue) walks. 
                Right: tottering (red) and no-tottering(blue) tree-patterns.\label{fig:totters} }
        \end{center}
\end{figure}

The DP algorithms of Sections \ref{sec:compute} and \ref{sec:untilN} enumerate balanced tree-patterns of order $h$ through the recursive extension of  balanced tree-patterns of order 2 defined by neighborhood matching sets of pairs of vertices.
According to Definition \ref{def:matching}, the whole sets of neighbors of a pair of vertices enter in the definition of their neighborhood matching sets. As a result, it can be the case in a tree-pattern that a vertex appears simultaneously as the parent and a child of a second vertex. This phenomenon is the tree counterpart of a phenomenon observed in the context of walk-based graph kernels, where a random walk under extension could return to a visited vertex just after leaving it. This behavior was called {\em tottering} in \cite{Mahe2005Graph}, and following this terminology, we refer to a tree-pattern in which a vertex appears simultaneously as the parent and a child of a second vertex as a {\em tottering tree-pattern}. Figure \ref{fig:totters} illustrates the tottering phenomenon.

In many cases these tree-patterns are likely to be uninformative features. In particular they are not proper subgraphs of the initial graphs. Even worse, the ratio of the number of tottering tree-patterns over the number of non-tottering tree-patterns quickly increases with the depth $h$ of the trees, suggesting that informative patterns corresponding to deep trees might be hidden by the profusion of tottering tree-patterns. In order to tackle this issue we now adapt an idea of \cite{Mahe2005Graph} to filter out these spurious tottering tree-patterns in the kernels presented in Sections \ref{sec:definition} and \ref{sec:gartner}. Tottering can be prevented by adding constraints in the tree-pattern counting function, according to the following definition.

\begin{definition}[No-tottering tree-pattern counting function]\label{def:nototter-counting}
From the tree-pattern counting function of Definition \ref{def:tree-counting}, a {\em no-tottering tree-pattern counting function} can be defined for the tree $t= (\mathcal{V}_t,\mathcal{E}_t)$, with $\mathcal{V}_t = (n_1,\dots,n_{|t|})$, and the graph  $G= (\mathcal{V}_G,\mathcal{E}_G)$, with $\mathcal{V}_G = (v_1,\dots,v_{|\mathcal{V}_G|})$, as
\begin{equation*}
\begin{split}
\psi^{NT}_t(G) = \big\vert \big\{ (\alpha_1,\dots,\alpha_{|t|})  \in [1,|\mathcal{V}_G|]^{|t|} : \quad & (v_{\alpha_1},\dots, v_{\alpha_{|t|}}) = \text{pattern}(t) \\
& \wedge  (n_{i}, n_{j}), (n_{j}, n_{k}) \in \mathcal{E}_t \iff \alpha_{i} \neq \alpha_{k} \big\}   \big\vert.
\end{split}
\end{equation*}
\end{definition}
Following Definition \ref{def:kernel}, a graph kernel based on no-tottering tree-patterns can be defined from this no-tottering tree-pattern counting function.
\begin{definition}[No-tottering tree-pattern kernel]\label{def:kernel-nototter}
A graph kernel $K^{NT}$ based on no-tottering tree-patterns is given for the graphs $G_1$ and $G_2$ by
\begin{equation}\label{eq:kernel-nototter}
K^{NT}(G_1,G_2) = \sum_{t \in \mathcal{T}} w(t) \psi_t^{NT}(G_1) \psi_t^{NT}(G_2),
\end{equation}
where $\mathcal{T}$ is a set of trees, $w : \mathcal{T} \rightarrow \mathbb{R}$ is a tree weighting functional and $\psi_t^{NT}$ is the no-tottering tree-pattern counting function of Definition \ref{def:nototter-counting}.
\end{definition}

This latter definition therefore extends the tree-pattern kernel of Definition \ref{def:kernel} to the no-tottering case.
However, due to the additional constraints on the set of acceptable patterns, the DP framework based on neighborhood matching set described in Sections \ref{sec:compute} and \ref{sec:untilN} does not hold any longer.
In \cite{Mahe2005Graph}, the following graph transformation was introduced in order to filter tottering walks.

\begin{definition}[Graph transformation]\label{def:transfo}
For a graph $G=(\mathcal{V}_G,\mathcal{E}_G)$, we let its  {\em transformed graph}  $G' = (\mathcal{V}_{G'},\mathcal{E}_{G'})$ be defined by:
\begin{itemize}
\item $ \mathcal{V}_{G'} =  \mathcal{V}_G \cup \mathcal{E}_G $,
\item $ \mathcal{E}_{G'} = \cbr{\br{v,\br{v,t}} | v \in \mathcal{V}_G, (v,t) \in \mathcal{E}_G } \; \cup \; \cbr{\br{\br{u,v},\br{v,t}} | \br{u,v},\br{v,t} \in \mathcal{E}_G, u \neq t} $,
\end{itemize}
and  labeled as follows:
\begin{itemize}
\item for a node $v' \in \mathcal{V}_{G'}$ the label is either $l(v') = l(v')$ if $v' \in \mathcal{V}_{G}$, or $l(v') = l(v)$ if $v' = (u,v) \in \mathcal{E}_{G}$,
\item for an edge $e' = (v'_1,v'_2)$ between two vertices $v'_1 \in \mathcal{V}_{G} \cup \mathcal{E}_{G} $ and $v'_2 \in \mathcal{E}_{G}$, the label is simply given by $l(e') = l(v'_2)$.
\end{itemize}
\end{definition}

\begin{figure}[h]
        \begin{center}
        \includegraphics[width=8cm]{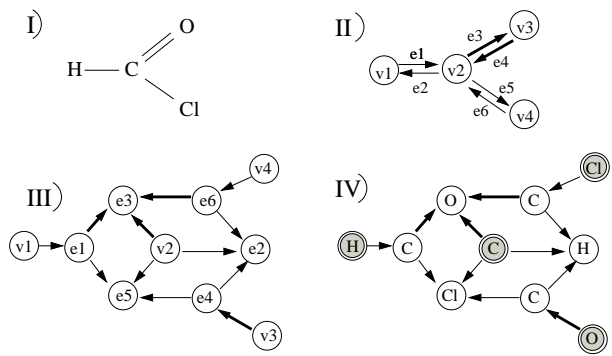}
        \caption{The graph transformation. I) The original molecule. II) The corresponding graph $G = (\mathcal{V}_G,\mathcal{E}_G)$.
III) The transformed graph. IV) The labels on the transformed graph. Note that different widths stand for
different edges labels, and gray nodes are the nodes belonging to $\mathcal{V}_G$.\label{fig:newgraph}}
        \end{center}
\end{figure}

This graph transformation is illustrated in Figure \ref{fig:newgraph} for the graph corresponding to the chemical compound of Figure \ref{fig:labeled}.
Based on this graph transformation, \cite{Mahe2005Graph} proved that there is a bijection between the set of no-tottering walks of a graph and the set of walks of its transformed graph that start on a vertex corresponding to a vertex of the original graph.
In a similar way, we show below that there is a bijection between the set of no-tottering tree-patterns found in a graph and the set of tree-patterns found in its transformed graph rooted in a vertex corresponding to a vertex of the original graph.
This is summarized in the following proposition, which proof is postponed in Appendix \ref{app:proof-nototter}.

\begin{proposition}\label{prop:nototter-kernel}
If we let $G'_1$ (resp. $G'_2$) be the transformed graph of $G_1$ (resp. $G_2)$, the no-tottering tree-pattern kernel of Definition \ref{def:kernel-nototter} is given by
\begin{align*}
K^{NT}(G_1,G_2) & = \sum_{t \in \mathcal{T}} w(t) \psi_t^{NT}(G_1) \psi_t^{NT}(G_2) \\
                             & = \sum_{t \in \mathcal{T}} w(t) \psi^{\{V_{G_1}\}}_t(G'_1) \psi^{\{V_{G_2}\}}_t(G'_2),
\end{align*}
where, if $G'$ is the transformed graph of $G$ given by Definition \ref{def:transfo},  $V_G \subset \mathcal{V}_{G'}$ is the set of vertices of $G'$ corresponding to the vertices of $G$, and $\displaystyle \psi_t^{\{v_1,\dots,v_n\}}(G) = \sum_{i=1}^n \psi_t^{(v_i)}(G)$.
\end{proposition}

This proposition shows that we can compute no-tottering extensions of the kernels of Definitions \ref{def:kernel-size} and \ref{def:kernel-branch}, and of the until-N kernel extension of Equation \ref{eq:kernel-untilN},  using the graph transformation of Definition \ref{def:transfo} and the original DP algorithms of Sections \ref{sec:compute} and \ref{sec:untilN}.
However, this operation comes at the expense of an increase in the cost of computing the kernel.
More precisely, by definition of the graph transformation, we have $|\mathcal{V}_{G'}| = |\mathcal{V}_G| + |\mathcal{E}_G|$.
Moreover, as noticed by \cite{Mahe2005Graph}, the maximum out-degree of the vertices of the transformed graph is equal to that of the original graph.
As a result, the worst case complexity of evaluating the functional $k_h(u,v)$ of Propositions \ref{prop:comput-size}, \ref{prop:comput-branch} and \ref{prop:untilN} is the same if $u$ and $v$ belong to $\mathcal{V}_{G'_1}$ and $\mathcal{V}_{G'_2}$, or  $\mathcal{V}_{G_1}$ and $\mathcal{V}_{G_2}$.
It follows that for the graphs $G_1$ and $G_2$ we have 
$$
\mathcal{O}\big(K^{NT}(G_1,G_2)\big) = \frac{(|\mathcal{V}_{G_1}|+|\mathcal{E}_{G_1}|) (|\mathcal{V}_{G_2}|+|\mathcal{E}_{G_2}|)}{|\mathcal{V}_{G_1}| |\mathcal{V}_{G_2}|} \quad \mathcal{O}\big(K(G_1,G_2)\big),
$$
where $K$ is one of the kernels given in Equations \ref{eq:kernel-size}, \ref{eq:kernel-branch} and \ref{eq:kernel-untilN}, and $K_{NT}$ is its no-tottering extension of Definition \ref{def:kernel-nototter}.

\section{Experiments}\label{sec:experiments}

We now turn to the experimental section. The problem we consider is a binary classification task consisting in discriminating toxic from non-toxic molecules. Our main goal is to assess the relevance of tree-patterns graph kernels over their walk-based counterparts for this type of chemical applications. To do so, recall from section \ref{sec:gartner-def} that in the proposed kernels, the influence of the tree-patterns is controlled by the parameter $\lambda$.
When $\lambda$ tends to zero, the kernels converge to kernels based on the count of common walks in the graphs \citep{Gartner2003graph}.
For increasing $\lambda$, tree-patterns of increasing complexity are taken into account with increasing weight in the kernels. One can therefore study the relevance of tree-patterns by studying how the performance of the kernels evolves with $\lambda > 0$, and checking whether it improves over their walk-based counterpart obtained for $\lambda=0$.

The first step towards this goal is to evaluate  the kernels of Definitions  \ref{def:kernel-size} and  \ref{def:kernel-branch}, and therefore the original formulation presented in \cite{Ramon2003Expressivity}.
In a second step, we want to validate the extensions to these  kernels proposed in sections \ref{sec:untilN} and \ref{sec:nototters}.
On the one hand we will compare the results obtained with the until-N extension of the branching-based kernel (\ref{eq:kernel-untilN}) to its initial formulation (\ref{eq:kernel-branch}), and on the other hand we will compare the results obtained with the no-tottering extensions (\ref{eq:kernel-nototter}) of the size-based, branching-based, and until-N branching-based kernels to their original formulations.
Because our interest here is to get insights about the behavior of the different kernels, we report experimental results for varying values of the parameters entering their definition, namely the order $h$ of the patterns, and the pattern weighting parameter $\lambda$. In real-world applications one should of course design a procedure to select the best parameters from the date.

The classification experiments described below were carried out with a support vector machine based on the
different kernels tested. Each kernel was implemented in C++ within the open-source ChemCpp toolbox, and we used the open-source Python machine learning package PyML\footnote{Available at \texttt{http://pyml.sourceforge.net}}  to perform SVM classification.
The SVM prediction is obtained by taking the sign of a score function. 
However, by varying this zero decision threshold, it is possible to compute the evolution of the true positive rate versus the false positive rate in a curve known as the Receiver Operating Characteristic (ROC) curve.
The area under this curve, known as AUC for Area Under the ROC Curve, is often considered to be a safer indicator of the quality of a classifier than its accuracy \citep{Fawcett2003ROC}, being 1 for an ideal classifier, and 0.5 for a random classifier.
The results presented below are averaged AUC values obtained for 10 repetitions of a 5-fold cross-validation process.
Within each cross-validation fold, the $"C"$ soft-margin parameter of the SVM was optimized over a grid ranging from $10^{-3}$ to $10^3$, using an internal cross-validation method implemented in PyML.

We considered two public datasets of chemical compounds in our experiments. Both gather results
of mutagenicity assays, and while the first one \citep{King1996Structure-activity} is a standard benchmark
for evaluating chemical compounds classification, the second one \citep{Helma2004Data} was
introduced more recently. 
The first dataset contains 188 chemical compounds tested for mutagenicity on {\itshape Salmonella typhimurium}.
The molecules of this dataset belong to the family of aromatic and hetero-aromatic nitro compounds, and they are split into two classes: 125 positive examples with high mutagenic activity (positive levels of log mutagenicity), and 63 negative examples with no or low mutagenic activity.
The second database considered consists of 684 compounds classified as mutagens or non-mutagens according to a test known as the \emph{Salmonella}/microsome assay. This dataset is well balanced with 341 mutagens compounds for 343 non-mutagens ones.
Note that although the biological property to be predicted is the same, the two datasets  are fundamentally different.
While \cite{King1996Structure-activity} focused on a particular family of molecules, this dataset involves a set of very diverse chemical compounds, qualified as \emph{noncongeneric} in the original paper.
To predict mutagenicity, the model therefore needs to solve different tasks : in the first case it has to detect subtle differences between homogeneous structures, while in the second case it must seek regular patterns within a set of structurally different molecules.

\subsection{First Dataset}

\begin{figure}[h]
\includegraphics[width=0.5\textwidth]{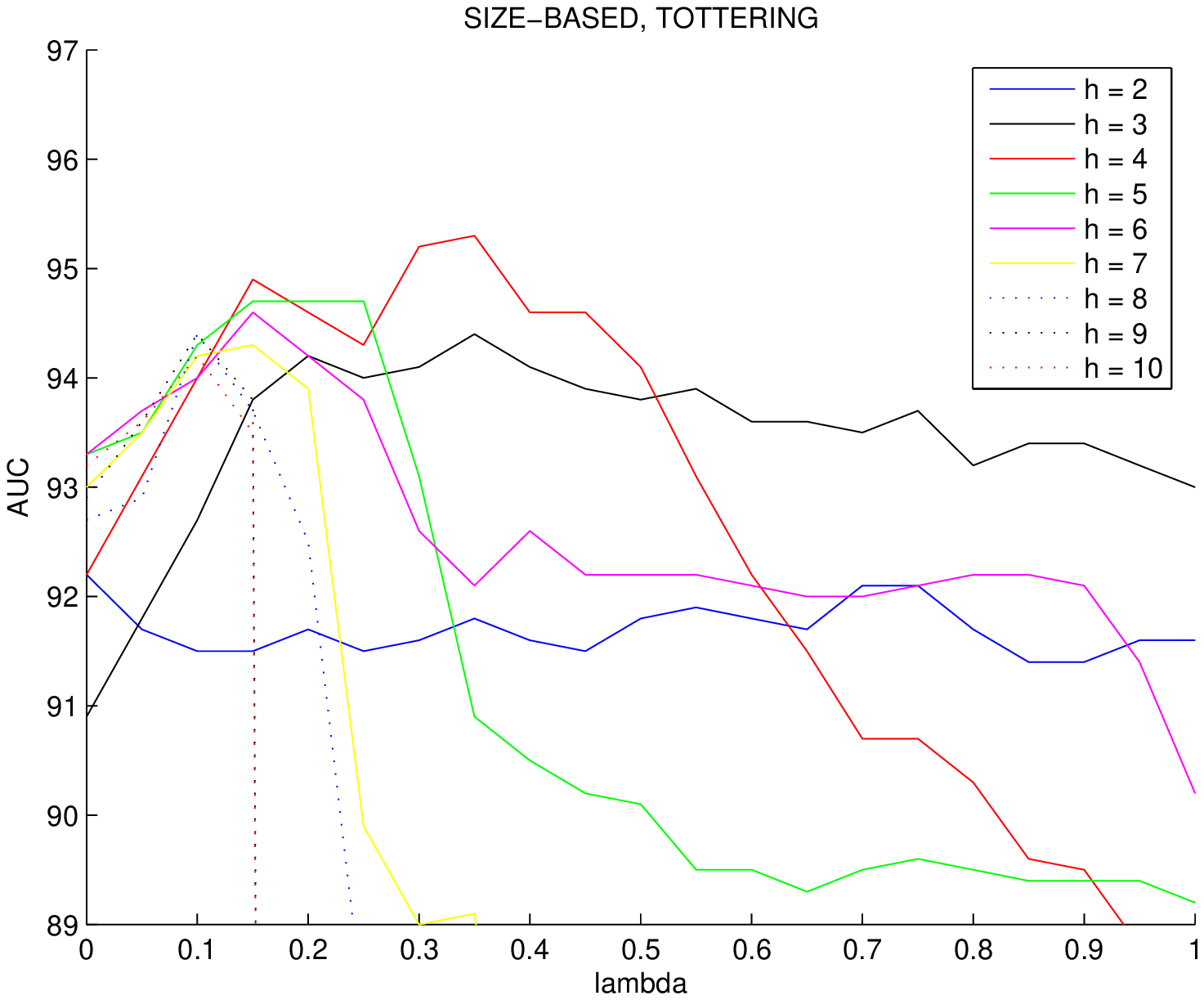}  
\includegraphics[width=0.5\textwidth]{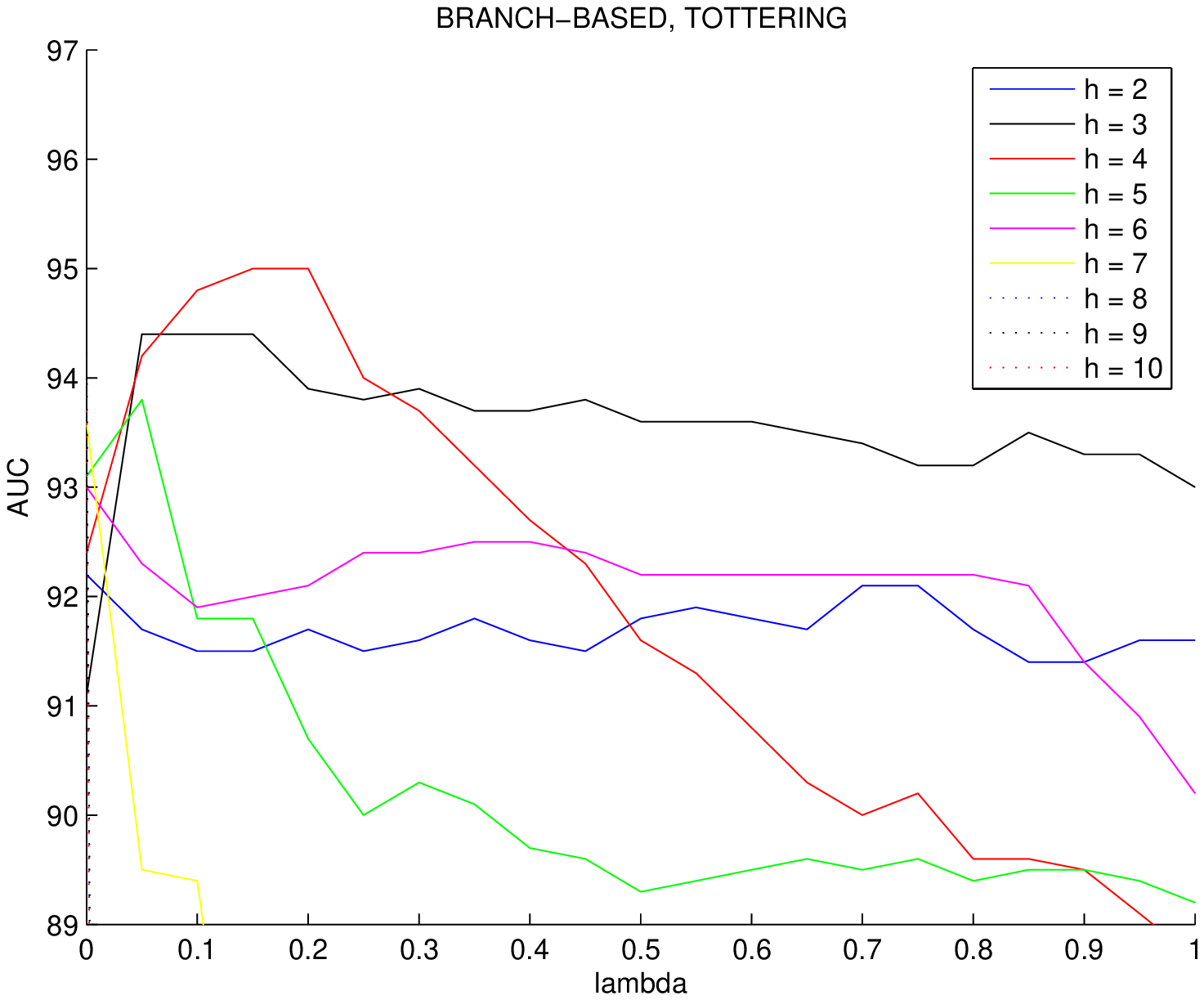}
\caption{First dataset. Evolution of the AUC with respect to $\lambda$ at different orders $h$.
Left: size-based kernel (\ref{eq:kernel-size}) ; Right: branching-based kernel (\ref{eq:kernel-branch}).}\label{fig:totters-mutag1}
\end{figure}

{\bf Tree-patterns Vs walk-patterns: }\\
Figure \ref{fig:totters-mutag1} shows the results obtained for the size-based (left) and branching-based (right) kernels of Definitions \ref{def:kernel-size} and \ref{def:kernel-branch}.
Each curve represents the evolution, for $0 \leq \lambda \leq 1$, of the AUC obtained from patterns of a given order $h$ taken between 2 and 10.

\begin{figure}[h]
\begin{center}
        \includegraphics[width=0.5\textwidth]{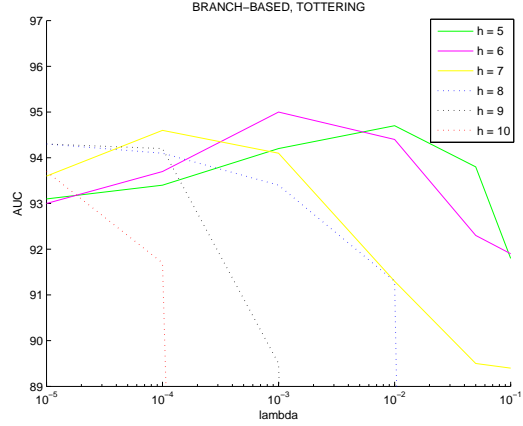} 
        \caption{First dataset, branching-based kernel (\ref{eq:kernel-branch}) . Evolution of the AUC at different orders $h$ for small values of $\lambda$.}\label{fig:totters-branch-small_mutag1}
\end{center}
\end{figure}

Because the corresponding AUC values start by increasing with $\lambda$, we can note from Figure \ref{fig:totters-mutag1} (left) that the introduction of tree-patterns is beneficial to the size-based kernel for patterns of order greater than two.
In the case of the branching-based kernel, Figure \ref{fig:totters-mutag1} (right) suggests that this is only true for patterns of order greater than 2 and smaller than 6, but Figure \ref{fig:totters-branch-small_mutag1} shows that, based on smaller values of $\lambda$, this is still the case for patterns up to order 7.
Taken together, Figures \ref{fig:totters-mutag1} and \ref{fig:totters-branch-small_mutag1}  show that the optimal AUC values obtained with the size- and branching-based kernels for patterns of order 2 to 7 are globally similar.
Interestingly however, the corresponding $\lambda$ values are systematically smaller in the case of the branching-based kernel.
This is due to the fact that, as noted in section \ref{sec:gartner-def}, the size-based penalization is stronger than the branching-based penalization. 
As a result, optimal $\lambda$ values observed using the size-based kernel are shifted towards zero using the branching-based kernel.

We can also note from Figures \ref{fig:totters-mutag1} and \ref{fig:totters-branch-small_mutag1} that optimal values of $\lambda$ tend to decrease for increasing $h$.
This is probably due to the fact that the number of tree-patterns increases exponentially with $h$, and, as a result, the kernels need to limit their individual influence.
Actually, we observe that higher order patterns, with $h > 7$, can  only be considered for sufficiently small values of $\lambda$.
For example, we note that the size-based kernel computation does not converge if we consider  patterns of order 10 and $\lambda$ greater than 0.15.
In the case of branching-based kernel, due to the weaker pattern penalization, this phenomenon is even emphasized, and in that case, $10^{-4}$ is the largest value acceptable for $\lambda$.
This difference in the way to penalize the patterns probably explains the fact that while a slight improvement over the walk-based kernel can be observed in the case of the size-based kernel when $h$ is greater than 7 (Figure \ref{fig:totters-mutag1}, left), the performance systematically decreases with the branching-based kernel (Figure \ref{fig:totters-branch-small_mutag1}).

Additionally, we note that because the size- and branching-based penalization of balanced trees of order 2 is the same, the results obtained for $h=2$ are identical with the two kernels. Surprisingly however, no improvement over the walk-based baseline is observed, which suggests that in this case, the tree-patterns do not bring additional information to that contained in the walk features, that consist here of simple pairs of connected atoms.

In conclusion, these experiments demonstrate the improvement of the tree-patterns graph kernels over their walk-based counterparts.
The impact of the tree-patterns is particularly marked for patterns of order 3 and 4, where the two kernels improve by more than 3\% the AUC of the corresponding walk-based kernel.
For patterns of increasing order, this figure gradually decreases, and for patterns of order greater than 7, it drops to 1 \% in the case of the size-based kernel, while no more improvement is observed with the branching-based kernel.
In both cases, optimal results are obtained for patterns of order 4, with AUC values of 95.3\% and 95.0\%. 
Finally, it is worth noting the combinatorial explosion in the number of patterns for large orders, which in practice limits the acceptable values of $\lambda$ to small values.\\


\begin{figure}[htbp]
\includegraphics[width=0.5\textwidth]{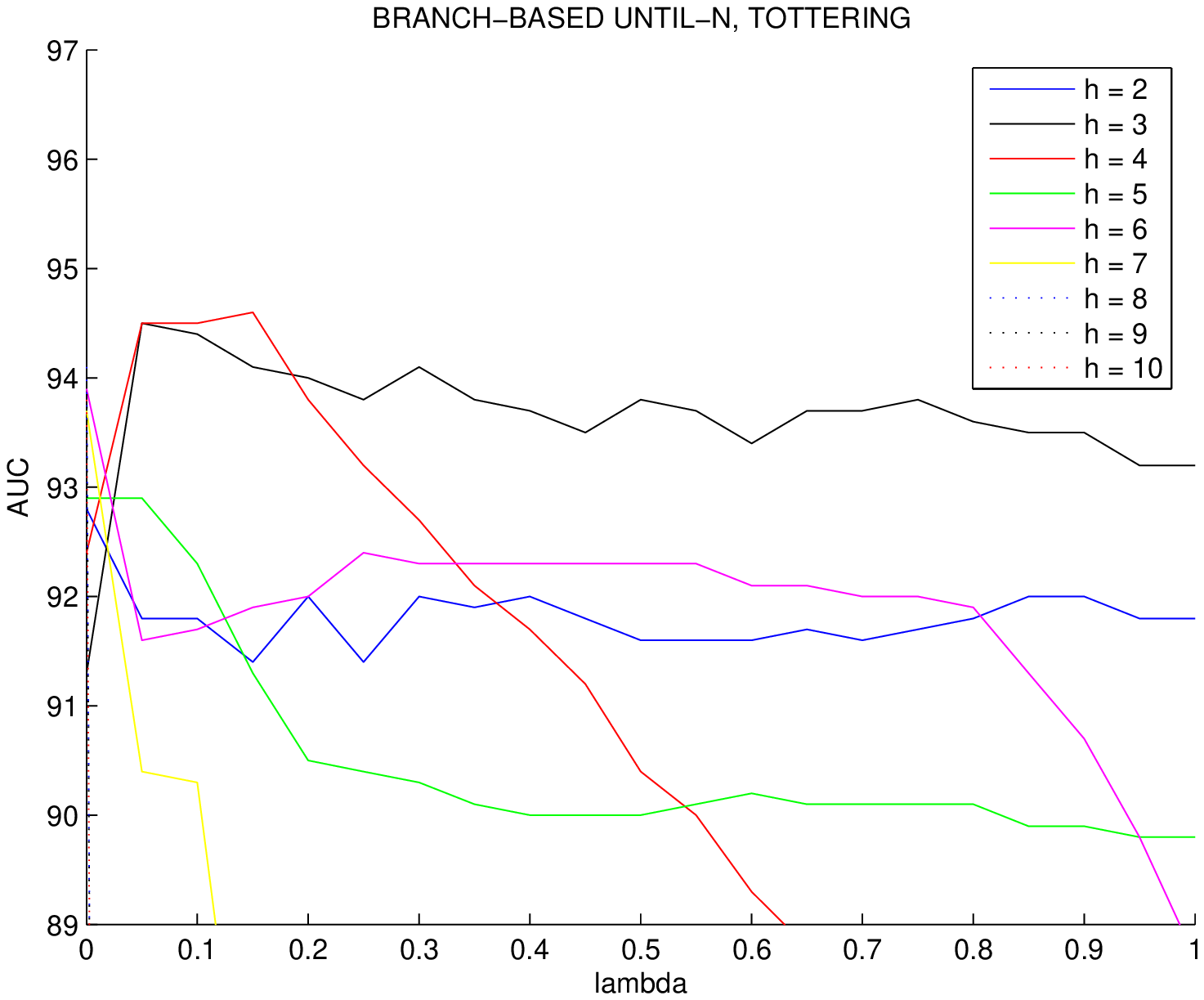} 
\includegraphics[width=0.5\textwidth]{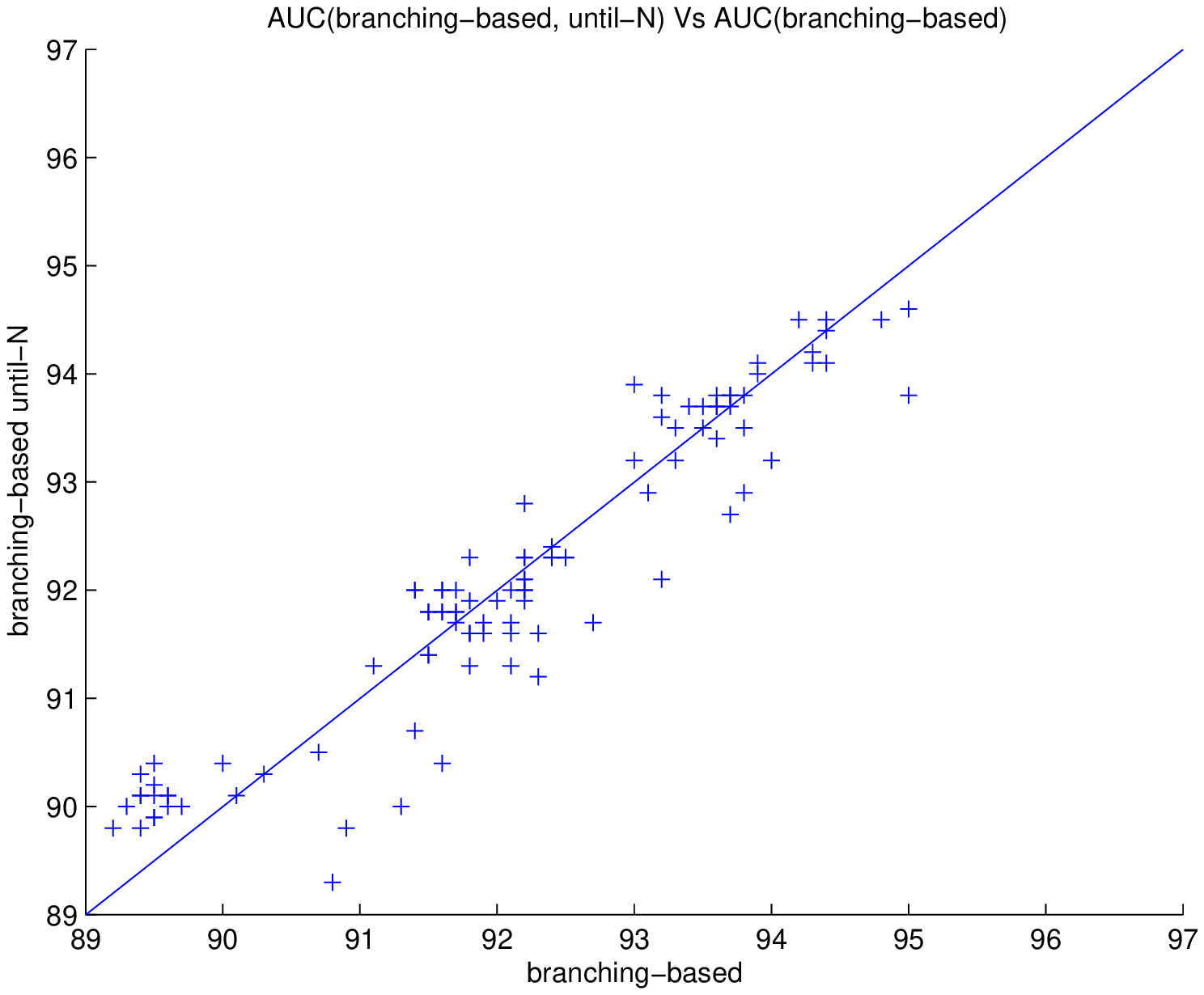} 
\caption{First dataset, until-N extension. Left: evolution of the AUC with respect to $\lambda$ at different orders $h$, for the until-N extension (\ref{eq:kernel-untilN}) of the branching-based kernel (\ref{eq:kernel-branch}). Right:  AUC values Vs original AUC values.}\label{fig:totters-untilN-mutag1}
\end{figure}

{\bf Until-N extension: }\\
Figure \ref{fig:totters-untilN-mutag1} presents the results of the until-N extension (\ref{eq:kernel-untilN}) of the branching-based kernel (\ref{eq:kernel-branch}).
The figure on the left-hand side, showing the evolution of the AUC for $2 \leq h \leq 10$ and $0 \leq \lambda \leq 1$, corresponds to that on the right-hand side of Figure \ref{fig:totters-mutag1}.
The figure on the right-hand side plots these AUC values versus corresponding values obtained using the original kernel (\ref{eq:kernel-branch}).

We can first notice strong similarities between the curve in the left-hand side and its original kernel counterpart.
This is confirmed in the right-hand curve where all the points lie near the diagonal line that represents the equivalence between the two kernels.
The fact that the differences between the two kernel formulations are barely noticeable is quite surprising since their associated feature spaces are intuitively quite different.
In section \ref{sec:untilN}, we mentioned that the feature space associated to the branching-based kernel is actually a subspace of the feature space associated to its until-N extension.
As a result, Figure \ref{fig:totters-untilN-mutag1} suggests that the extra features related to the until-N extension do not bear additional information into the kernel.
This hypothesis seems to be confirmed by the fact that the differences between corresponding walk-based kernels, observed for $\lambda = 0$, are not significant neither.
This might be explained by the fact that the dimensions of the corresponding feature space are probably strongly correlated due to the relation of inclusion existing between  trees and walks patterns of orders $n$, and those of order $n+1$. Another possible explanation for the lack of improvement of the until-N extension lies of course in the difficulty of learning in high dimension, suggesting that discriminating patterns of a given order are lost within the flood of patterns of greater orders taken into account by this until-N extension.

\begin{figure}[htbp]
\includegraphics[width=0.5\textwidth]{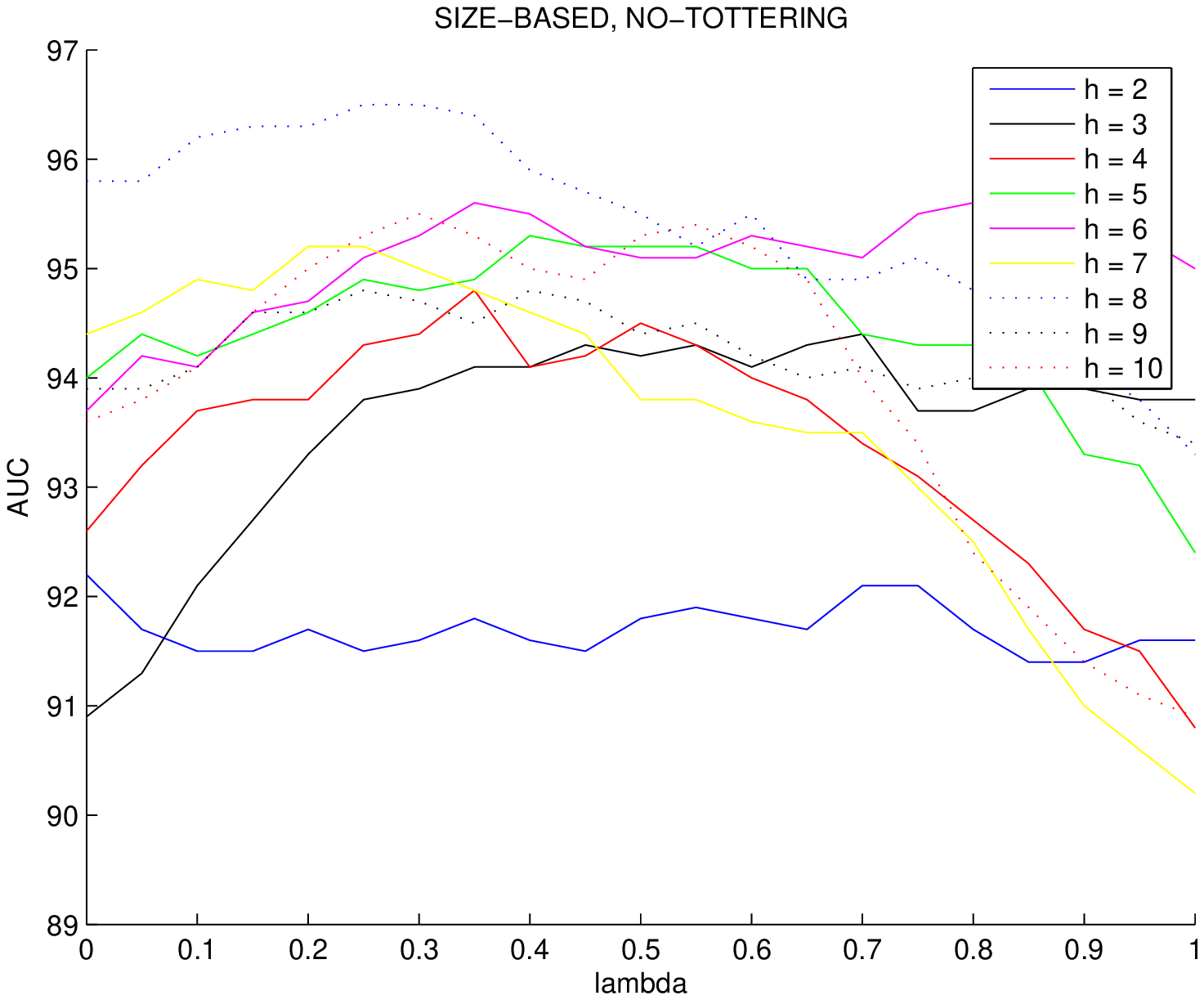}
\includegraphics[width=0.5\textwidth]{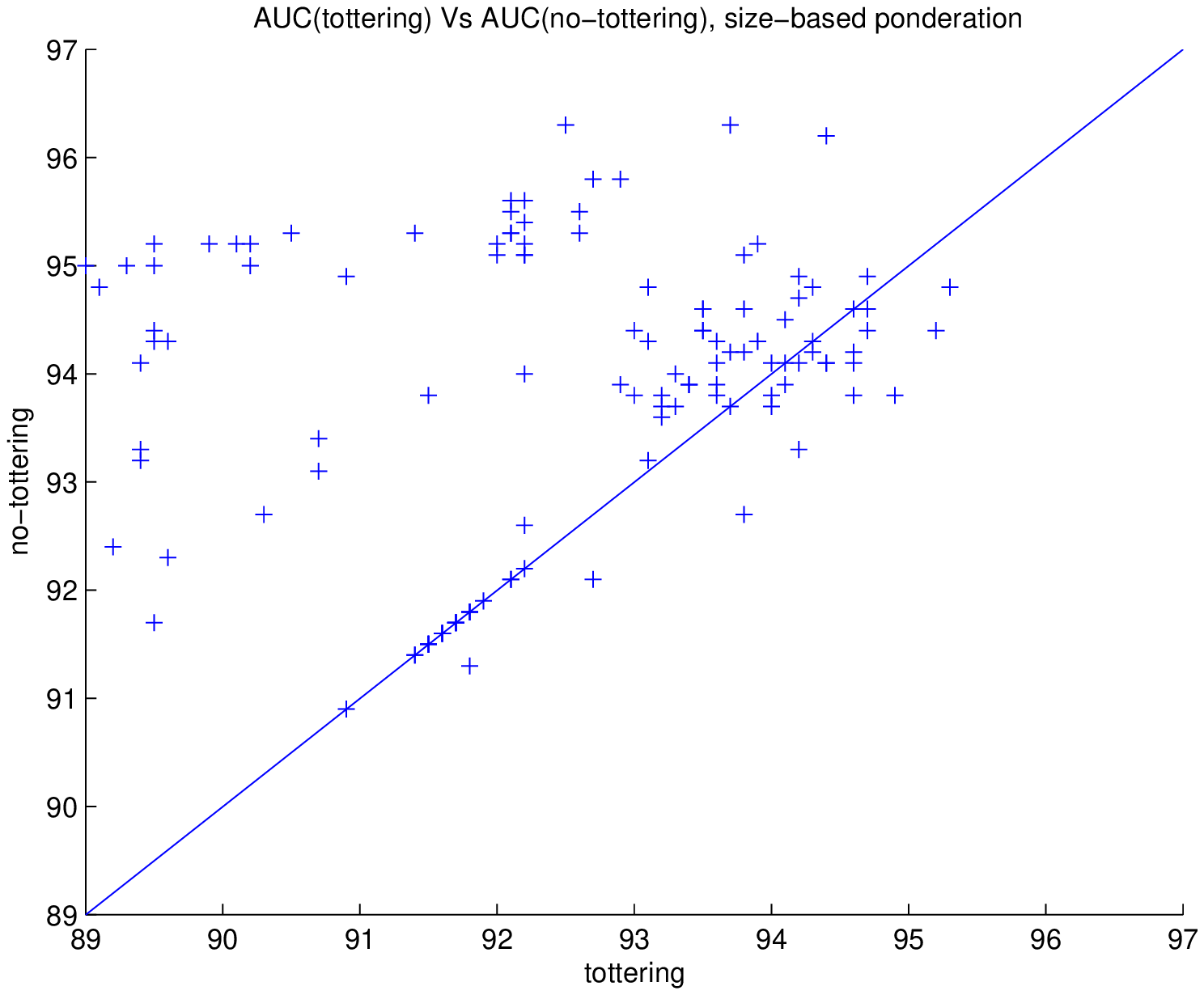}
\caption{First dataset. Left: evolution of the AUC with respect to $\lambda$ at different orders $h$ for the no-tottering extension (\ref{eq:kernel-nototter}) of the size-based kernel (\ref{eq:kernel-size}). Right: no-tottering AUC values Vs original AUC values.}\label{fig:nototters-size-mutag1}
\end{figure}

\begin{figure}[htbp]
\includegraphics[width=0.5\textwidth]{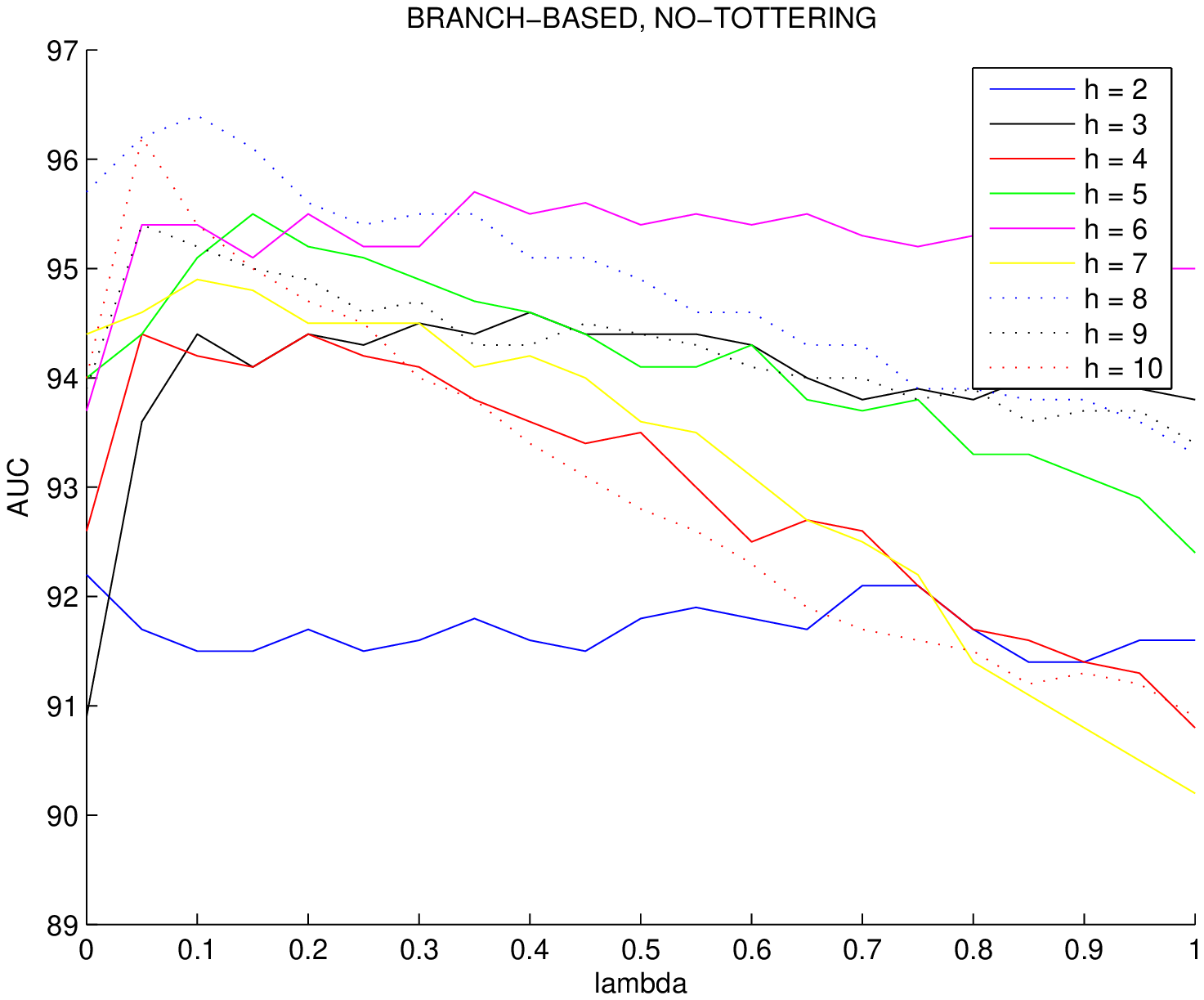}
\includegraphics[width=0.5\textwidth]{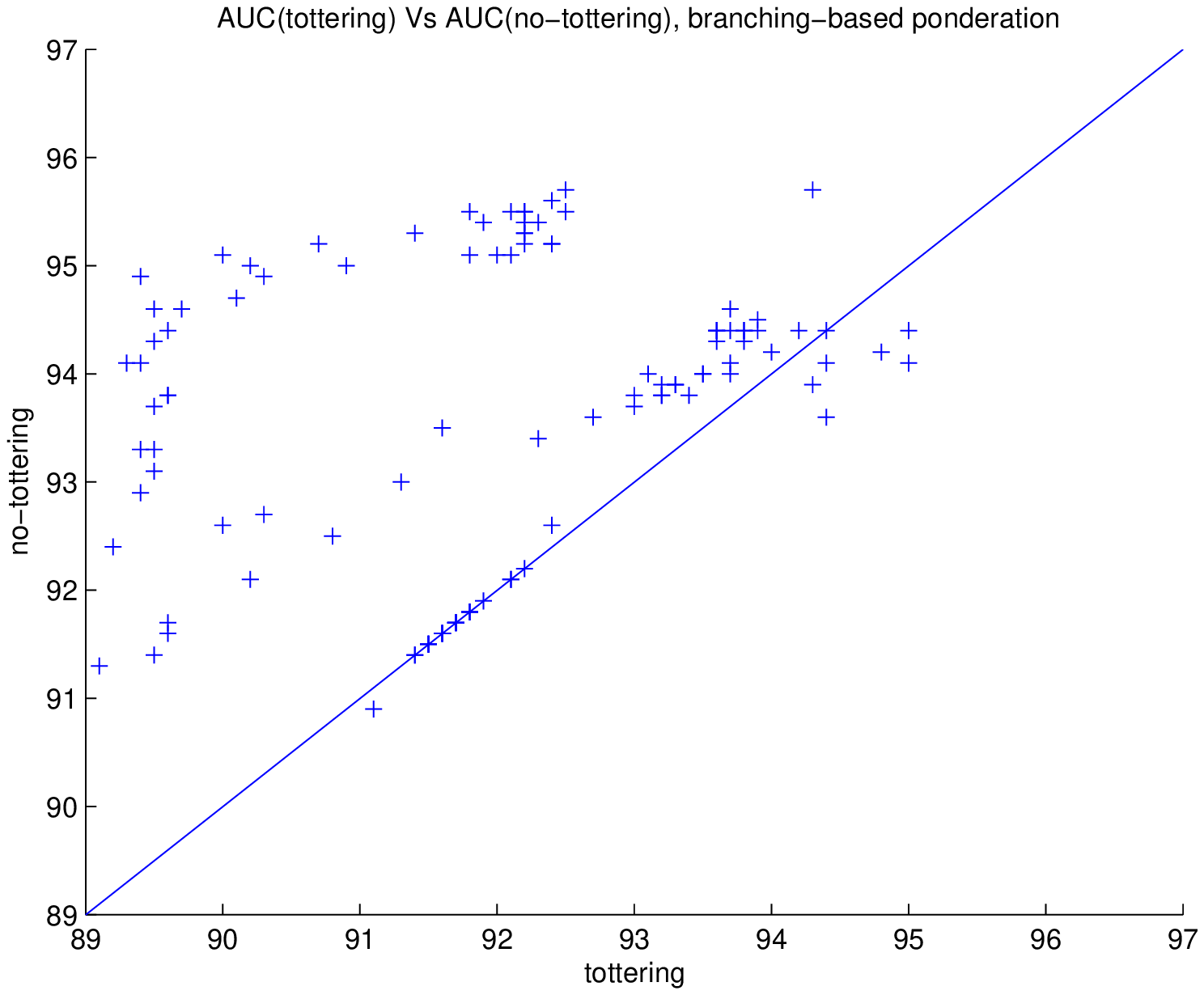} 
\caption{First dataset. Left: evolution of the AUC with respect to $\lambda$ at different orders $h$ for the no-tottering extension (\ref{eq:kernel-nototter}) of the branching-based kernel (\ref{eq:kernel-branch}). Right: no-tottering AUC values Vs original AUC values.}\label{fig:nototters-branch-mutag1}
\end{figure}

\begin{figure}[htbp]
\includegraphics[width=0.5\textwidth]{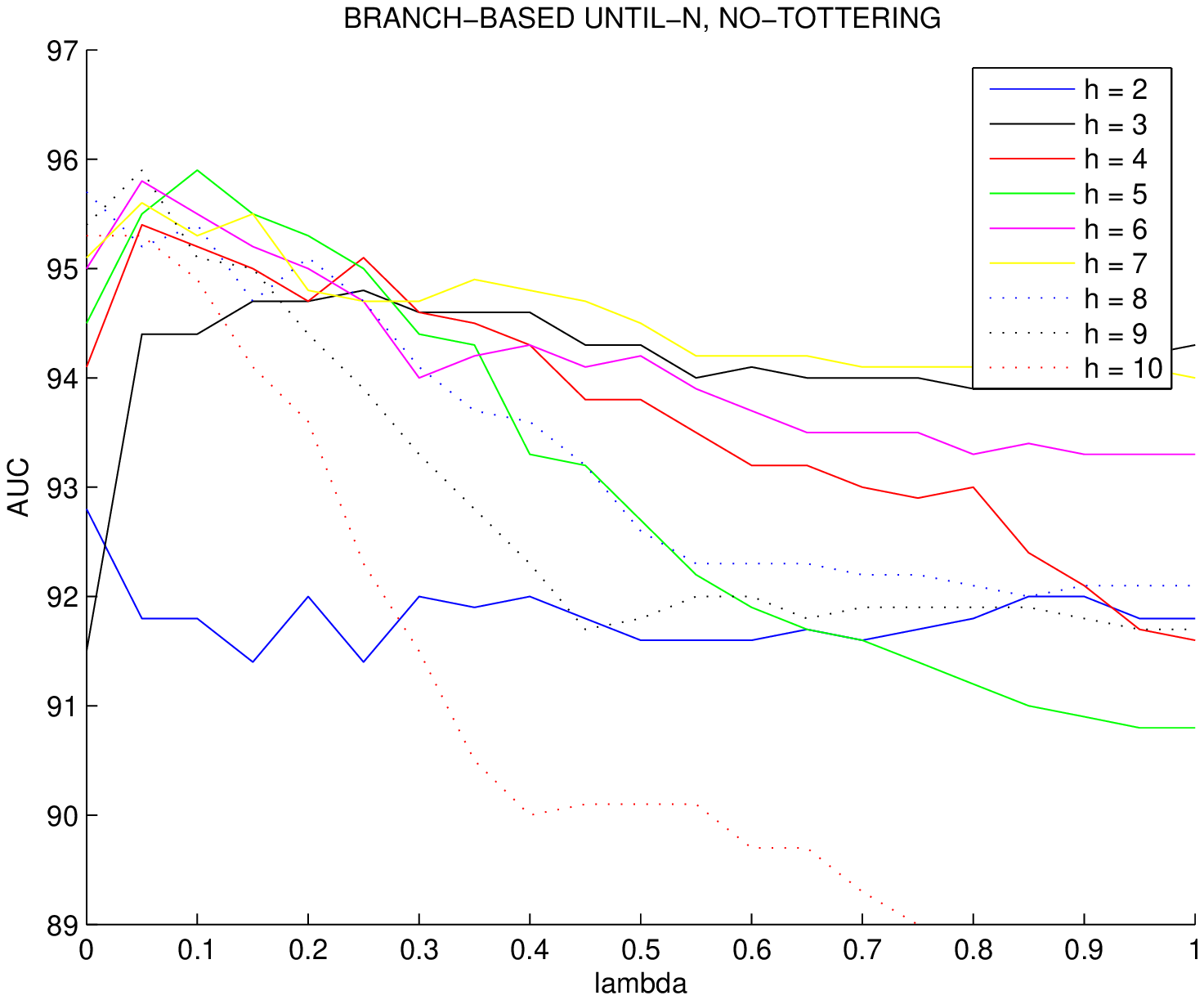}
\includegraphics[width=0.5\textwidth]{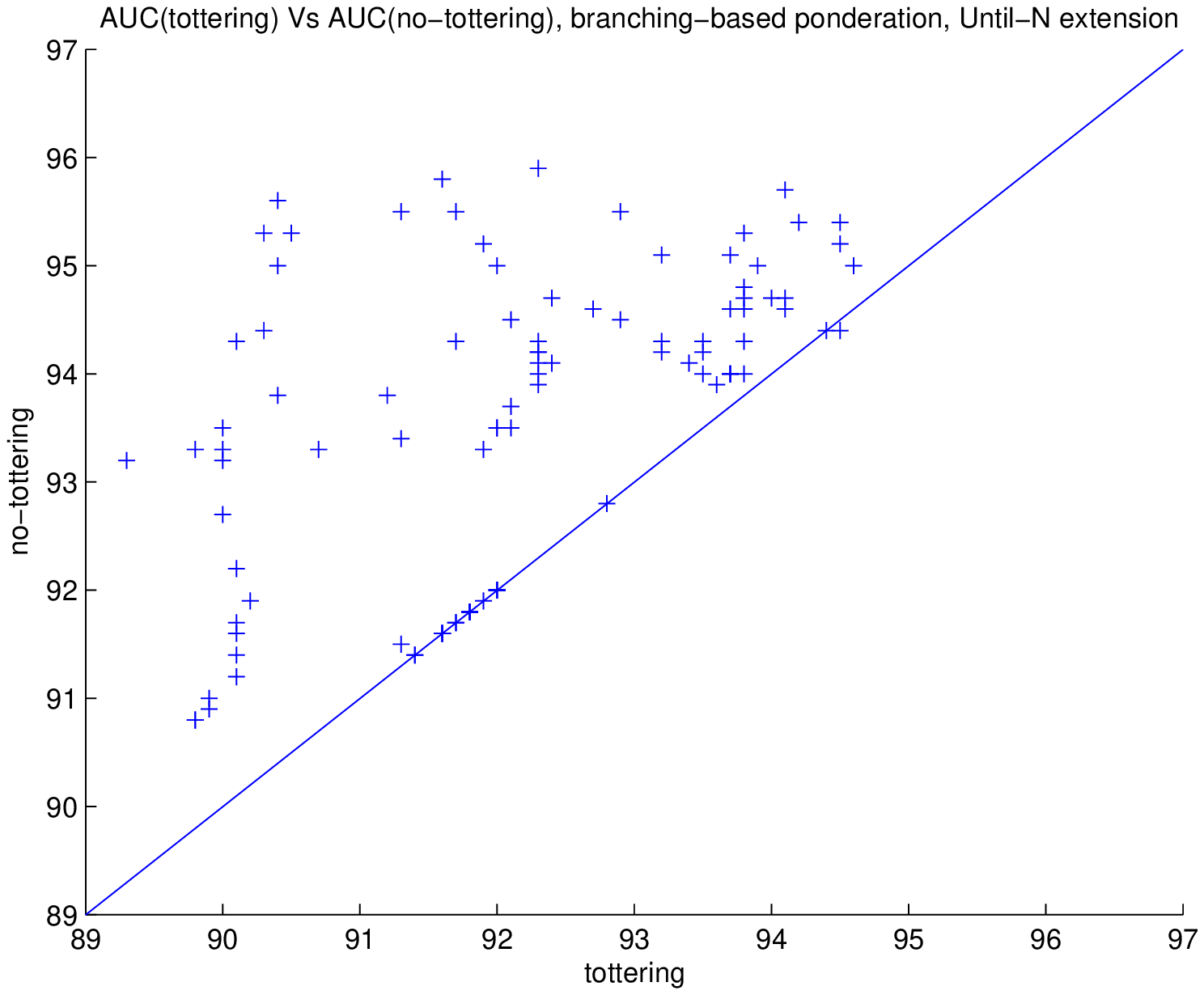}
\caption{First dataset. Left: evolution of the AUC with respect to $\lambda$ at different orders $h$ for the no-tottering extension (\ref{eq:kernel-nototter}) of the until-N branching-based kernel (\ref{eq:kernel-untilN}). Right: no-tottering AUC values Vs original AUC values.}\label{fig:nototters-branchUntilN-mutag1}
\end{figure}

{\bf No-tottering extension: }\\
Figures \ref{fig:nototters-size-mutag1}, \ref{fig:nototters-branch-mutag1} and \ref{fig:nototters-branchUntilN-mutag1} respectively show the results of the no-tottering extension (\ref{eq:kernel-nototter}) of the size-based (\ref{eq:kernel-size}), branching-based (\ref{eq:kernel-branch}), and until-N branching-based kernels (\ref{eq:kernel-untilN}).
The curves on the left-hand side show the evolution of AUC for  $2 \leq h \leq 10$ and $0 \leq \lambda \leq 1$, and the curves on the right-hand side plot these AUC values versus corresponding values obtained using the original kernels.

If we compare the results of the no-tottering extensions of the size-based and branching-based kernels (Figures \ref{fig:nototters-size-mutag1} and \ref{fig:nototters-branch-mutag1}), we can first note that the the introduction of tree-patterns is now systematically beneficial for $h > 2$ in both cases.
Moreover, we note that the kernel computations remain feasible for $h=10$ and $\lambda=1$, which means that the no-tottering extension limits the combinatorial explosion we observed with the original formulation.
While optimal results were obtained for $h=4$ using the original kernels, we observe that here, in both cases, the performance gradually increases from $h=3$ to an optimum value obtained for $h=8$.
At a given order, we note that the optimal AUC values obtained with the two kernels are similar, and that the corresponding $\lambda$ value is smaller in the case of the branching-based kernel, which is consistent with the observations made in the previous section.
Optimal AUC values are close to 96.5\% and improve over the values around 95\% observed with the initial formulation.
Importantly, we note that these optimal values are obtained using parametrizations of the kernels that lead to a combinatorial explosion in their initial formulation.
Finally, from the fact that almost all points lie above the diagonal in the right-hand curves, we can draw the conclusion that the no-tottering extension has almost consistently a positive influence on the classification in both cases.
It is worth noting however that, even though the introduction of no-tottering tree-patterns was shown to be beneficial, part of the overall improvement over their tottering counterparts is due to the no-tottering extension itself, since no-tottering walk-based kernels, observed for $\lambda=0$, already improve significantly over their tottering counterparts, especially for high order patterns.

We now turn to Figure \ref{fig:nototters-branchUntilN-mutag1} and the no-tottering extension  (\ref{eq:kernel-nototter}) of the until-N branching-based kernel (\ref{eq:kernel-untilN}).
We can first notice that conclusions similar to those related to the no-tottering extension of the branching-based kernel can be drawn: an improvement over the corresponding walk-based kernel is systematically observed for tree-patterns of order greater than 2, the kernel behaves more nicely (no combinatorial explosion), and the no-tottering extension consistently improves over the initial until-N branching-based kernel (right-hand curve).
Interestingly however, we note that optimal results obtained for $4 \leq h \leq 10$ tend to converge to an optimal value around 95.5\% (between 95.3 and 95.9\%) for a $\lambda$ value around 0.05.
While this global optimum is not as good as the overall optimal result obtained with the no-tottering branch-based kernel (Figure \ref{fig:nototters-branch-mutag1}), it still remains competitive (95.5\% Vs 96.5\%).
This observation contrasts with the the results obtained with the until-N extension in the tottering case, where patterns of a given order seemed to be lost in the amount of patterns of greater orders taken into account by the kernel.
This is due to the fact the the no-tottering extension limits the number of patterns to be detected, and suggests that patterns of different orders can now be considered simultaneously in the kernel.
This fact therefore suggests that in the no-tottering case, the until-N extension can help solving the problem of pattern order selection by taking a maximal pattern order large enough (here, $h > 4$).

\subsection{Second Dataset}


In this section, we apply the same analysis to the second dataset.\\

\begin{figure}[h]
\includegraphics[width=0.5\textwidth]{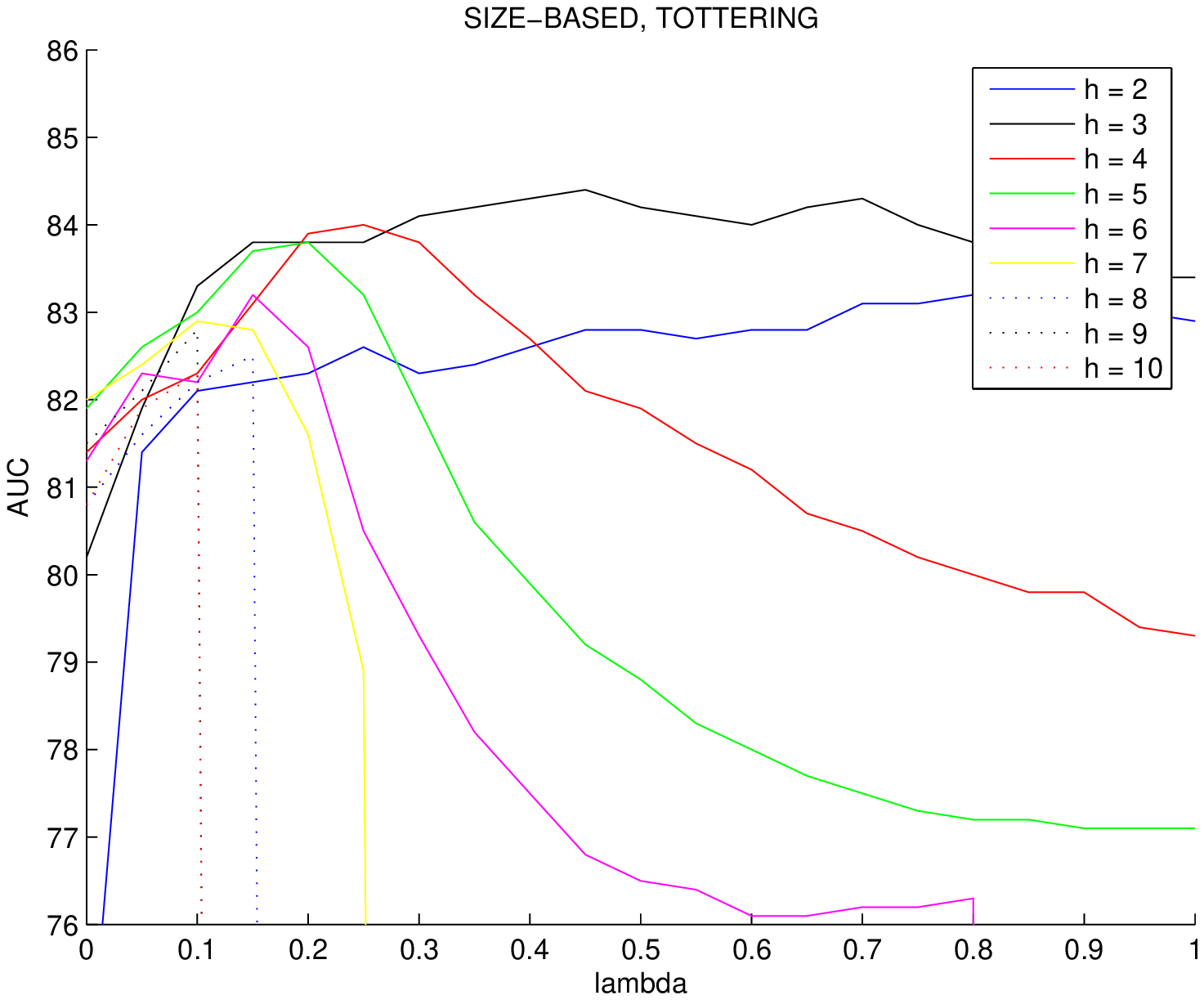}
\includegraphics[width=0.5\textwidth]{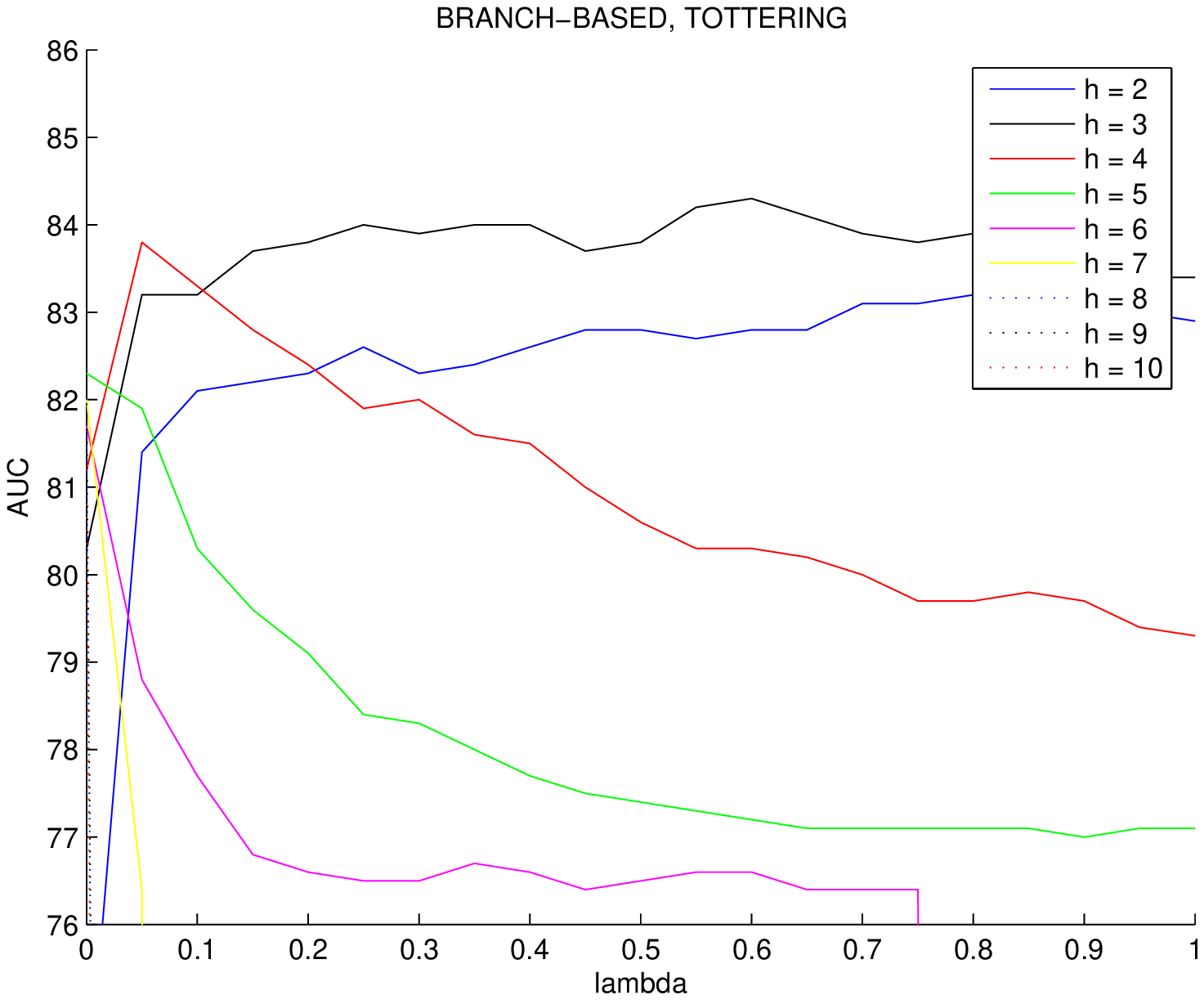}
\caption{Second dataset. Evolution of the AUC with respect to $\lambda$ at different orders $h$.
Left: size-based kernel (\ref{eq:kernel-size}) ; Right: branching-based kernel (\ref{eq:kernel-branch}).}\label{fig:totters-mutag2}
\end{figure}

{\bf Tree-patterns Vs walk-patterns: }\\
Figure \ref{fig:totters-mutag2} shows the results obtained with the original size-based (\ref{eq:kernel-size}) and branching-based (\ref{eq:kernel-branch}) kernels.
Several observations are consistent with those we drew with the fist dataset.
First, the introduction of tree-patterns has in both cases a positive influence on the classification, and is particularly marked for patterns of limited order (up to a relative improvement of 12\% for $h=2$, and 4.5\% for $h=3$).
Moreover, optimal values of the  $\lambda$ parameter are smaller in the case of the branching-based kernel, they decrease for increasing $h$, and quickly lead to a combinatorial explosion for high-order patterns.
Finally, we note that, in both cases, optimal AUC values are around 84\%, and are obtained for patterns of order 3 and 4, which is similar to the optimal order observed for the first dataset.
However, we can note the interesting difference that here, tree-patterns of order 2 improve dramatically the results over their walk counterparts, which suggests that different molecular features are to be detected in both datasets.\\

\begin{figure}[htbp]
\includegraphics[width=0.5\textwidth]{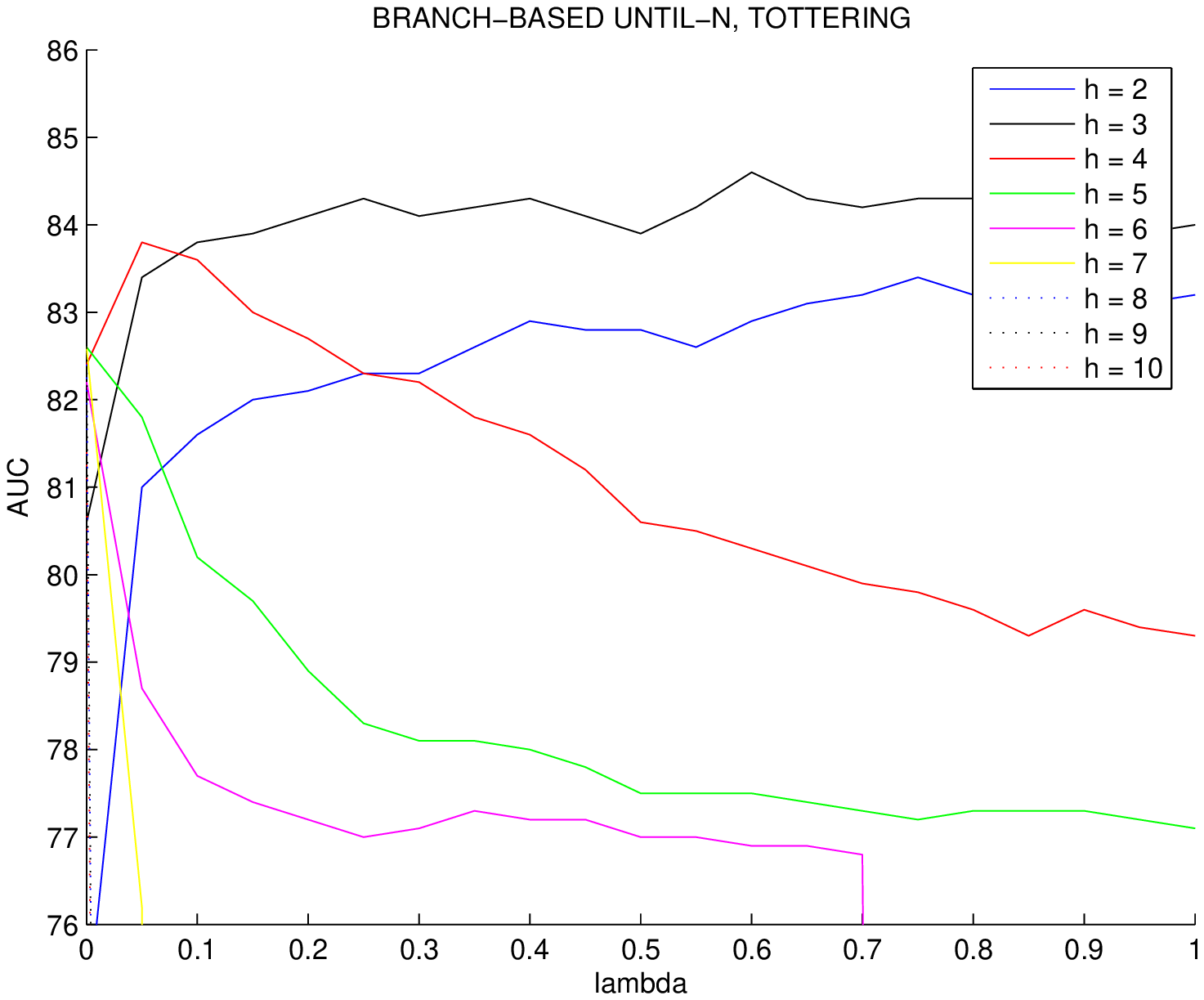}
\includegraphics[width=0.5\textwidth]{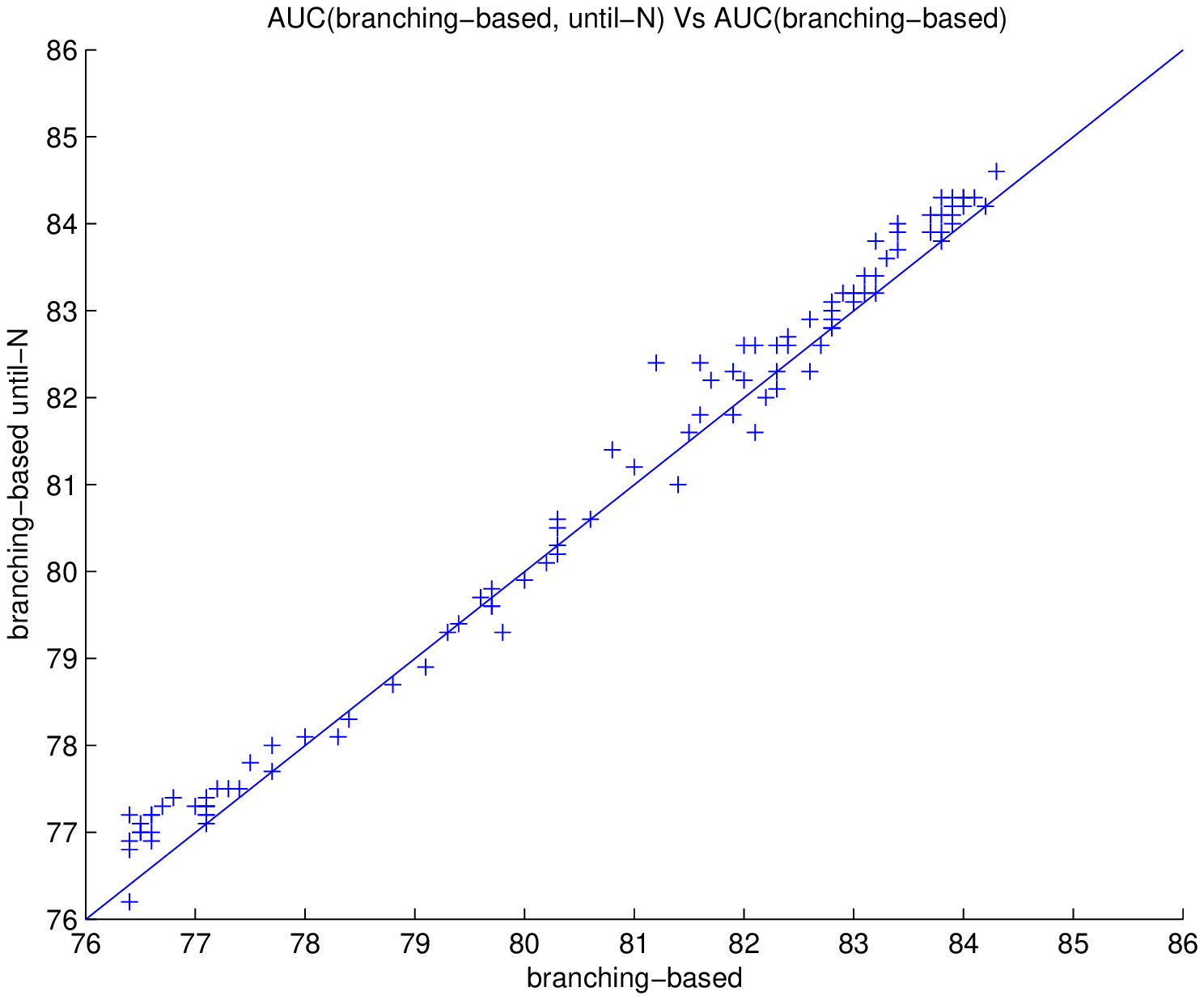} 
\caption{Second dataset, until-N extension. Left: evolution of the AUC with respect to $\lambda$ at different orders $h$ for the until-N extension (\ref{eq:kernel-untilN}) of the branching-based kernel (\ref{eq:kernel-branch}). Right: AUC values Vs original AUC values.}\label{fig:totters-untilN-mutag2}
\end{figure}

{\bf Until-N extension: }\\
Figure \ref{fig:totters-untilN-mutag2} shows the results obtained with the until-N extension  (\ref{eq:kernel-untilN}) of the branching-based kernel (\ref{eq:kernel-branch}).
Here again, observations are consistent with the first dataset.
In particular, we can note that the results obtained with and without the until-N extension are very similar, and this fact is even more pronounced here.
This second evidence confirms that the until-N extension is of little use in the original formulation of the kernel, most probably because patterns of a given order are drowned within the amount of patterns of greater orders.\\

\begin{figure}[h]
\includegraphics[width=0.5\textwidth]{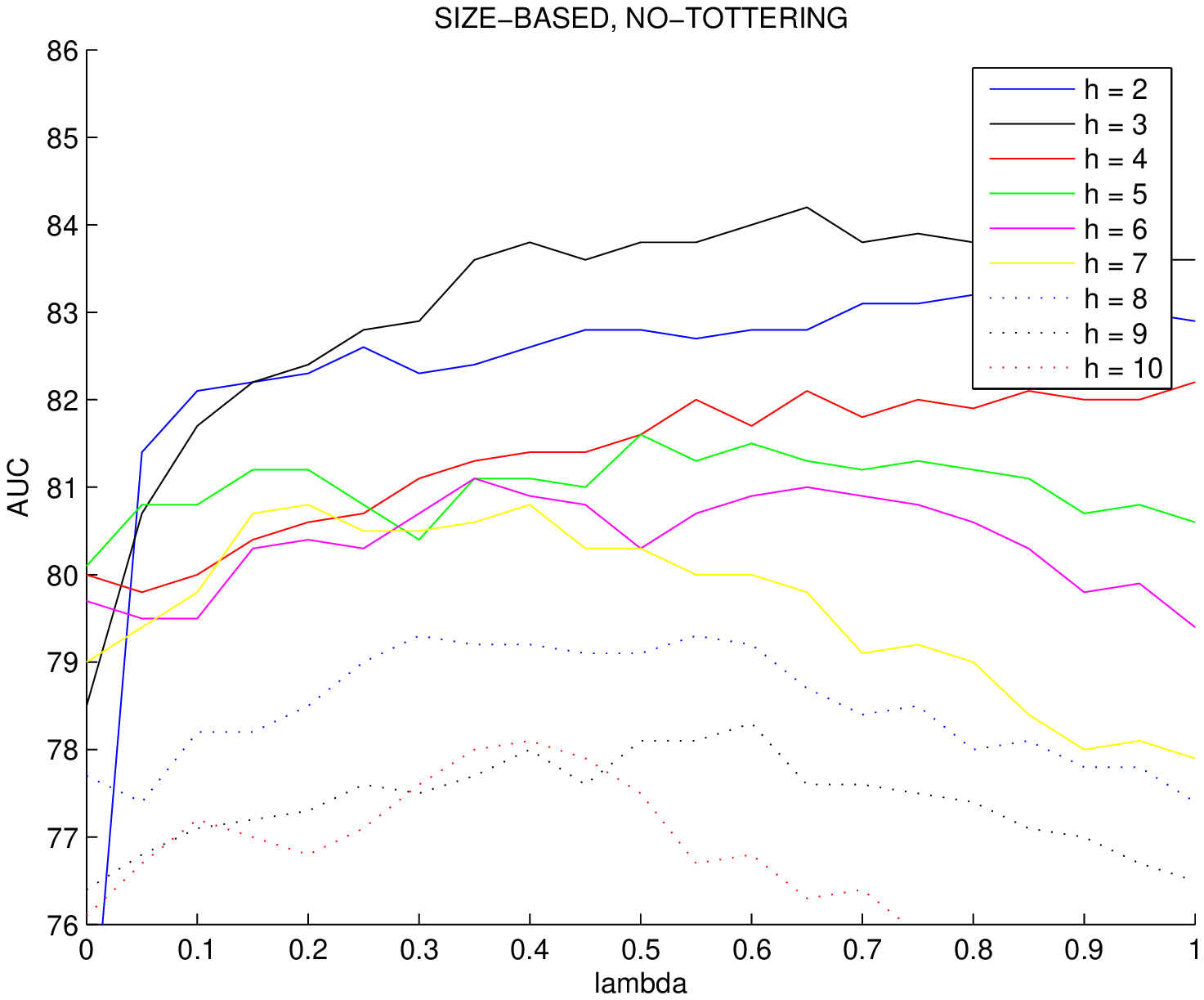}
\includegraphics[width=0.5\textwidth]{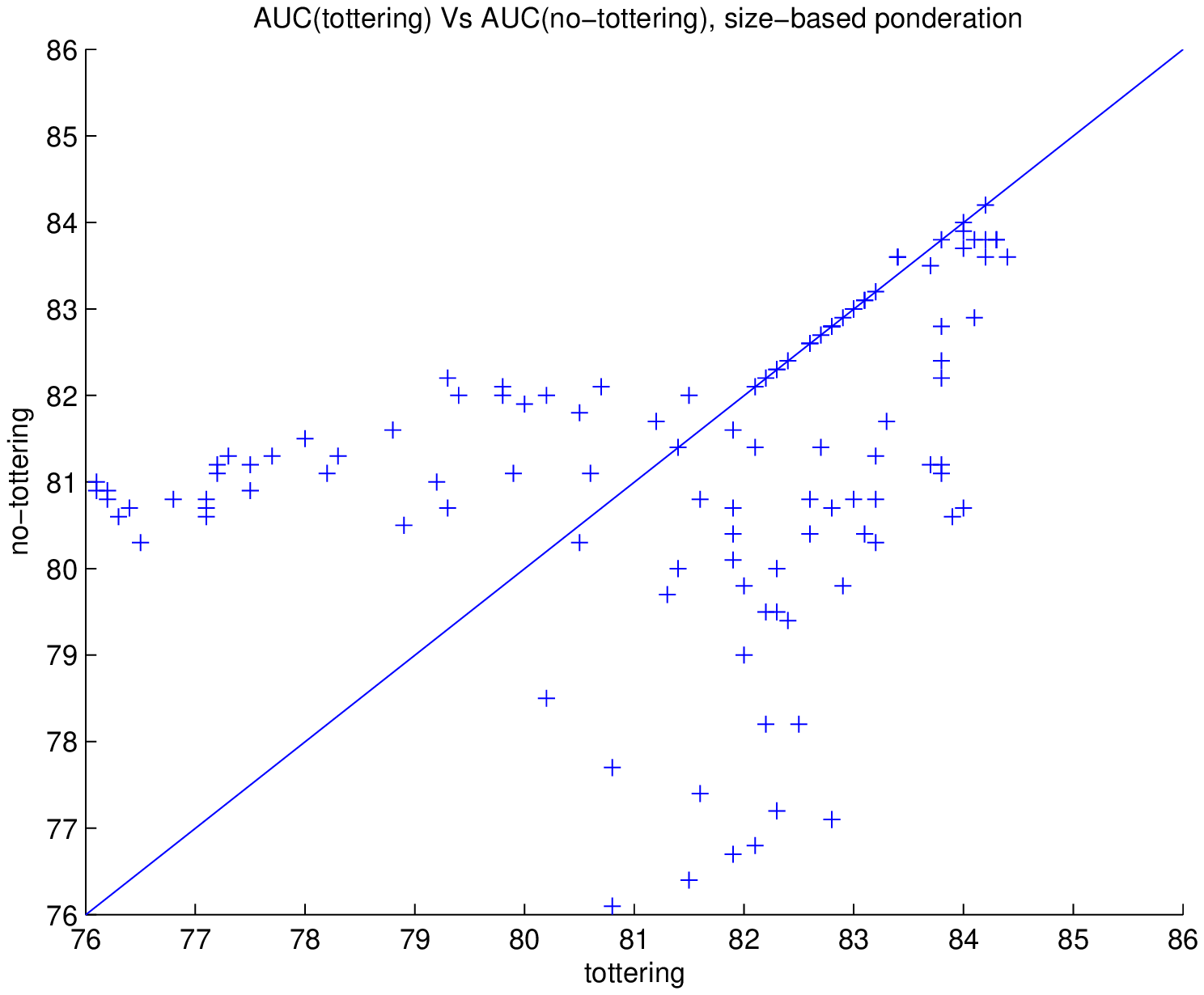}
\caption{Second dataset. Left: evolution of  the AUC with respect to $\lambda$ at different orders $h$ for the no-tottering extension (\ref{eq:kernel-nototter}) of the size-based kernel (\ref{eq:kernel-size}). Right: no-tottering AUC values Vs original values.}\label{fig:nototters-size-mutag2}
\end{figure}

\begin{figure}[h]
\includegraphics[width=0.5\textwidth]{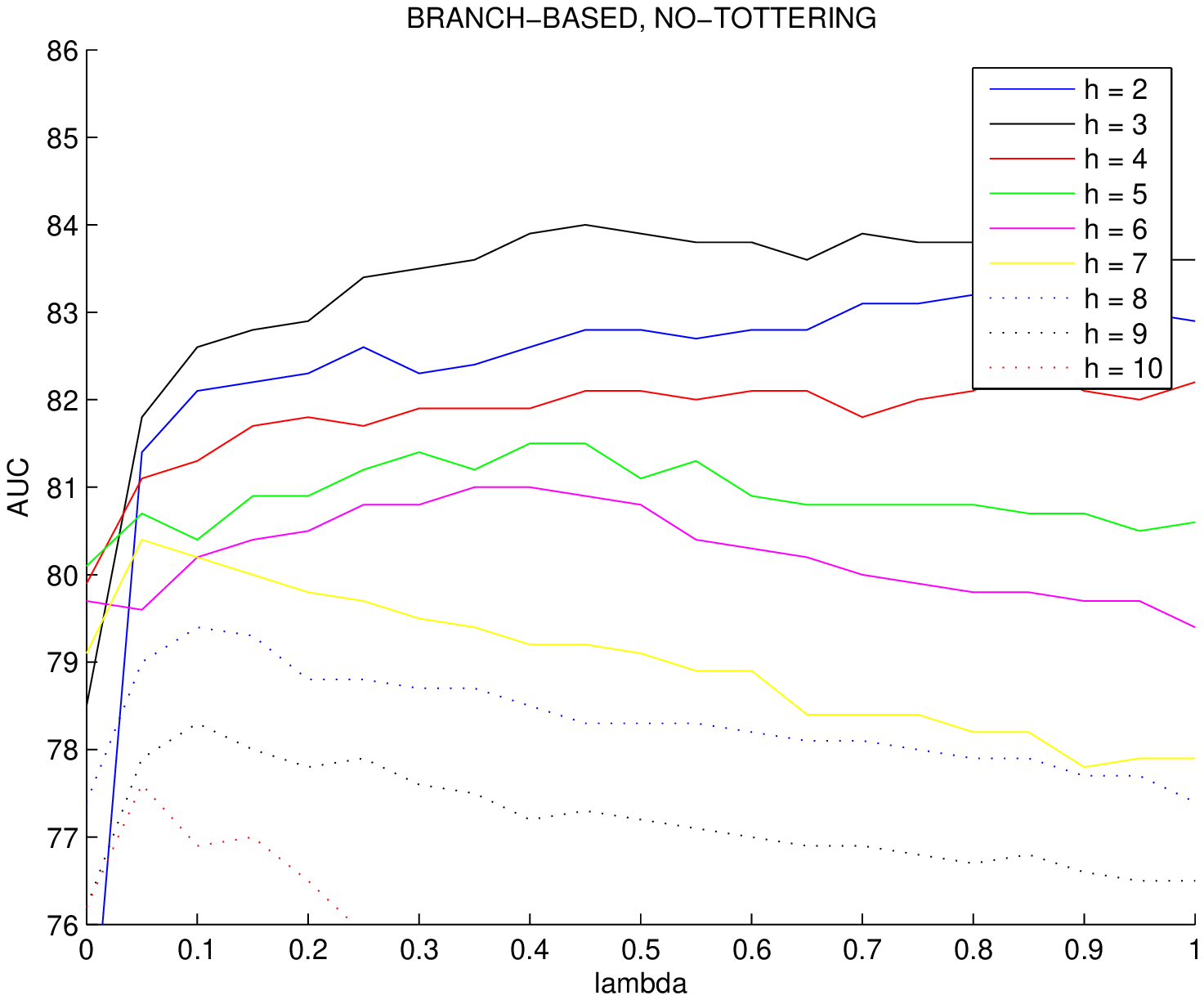}
\includegraphics[width=0.5\textwidth]{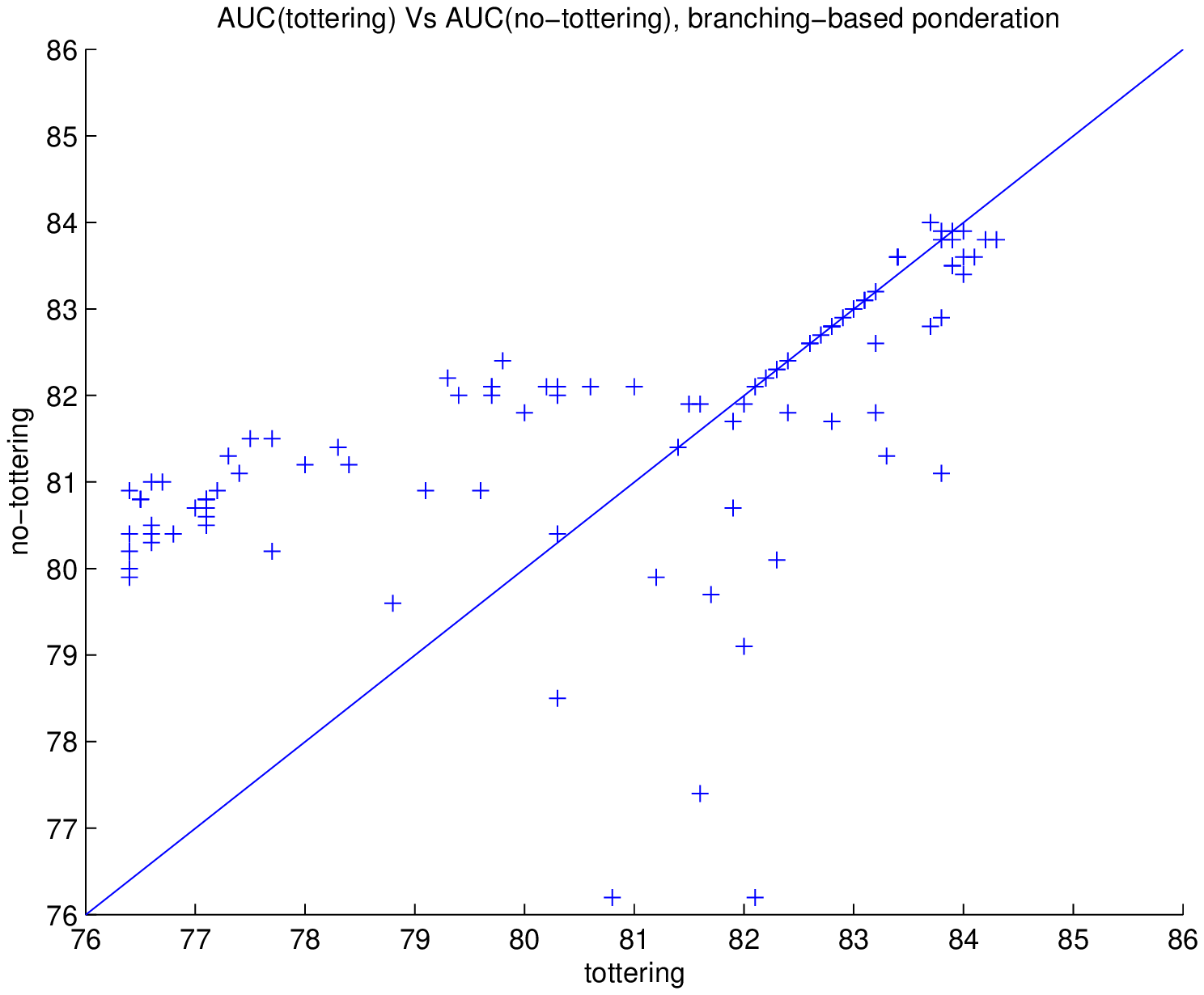} 
\caption{Second dataset. Left: evolution of the AUC with respect to $\lambda$ at different orders $h$ for the no-tottering extension (\ref{eq:kernel-nototter}) of the branching-based kernel (\ref{eq:kernel-branch}). Right: no-tottering AUC values Vs original values.}\label{fig:nototters-branch-mutag2}
\end{figure}

\begin{figure}[h]
\includegraphics[width=0.5\textwidth]{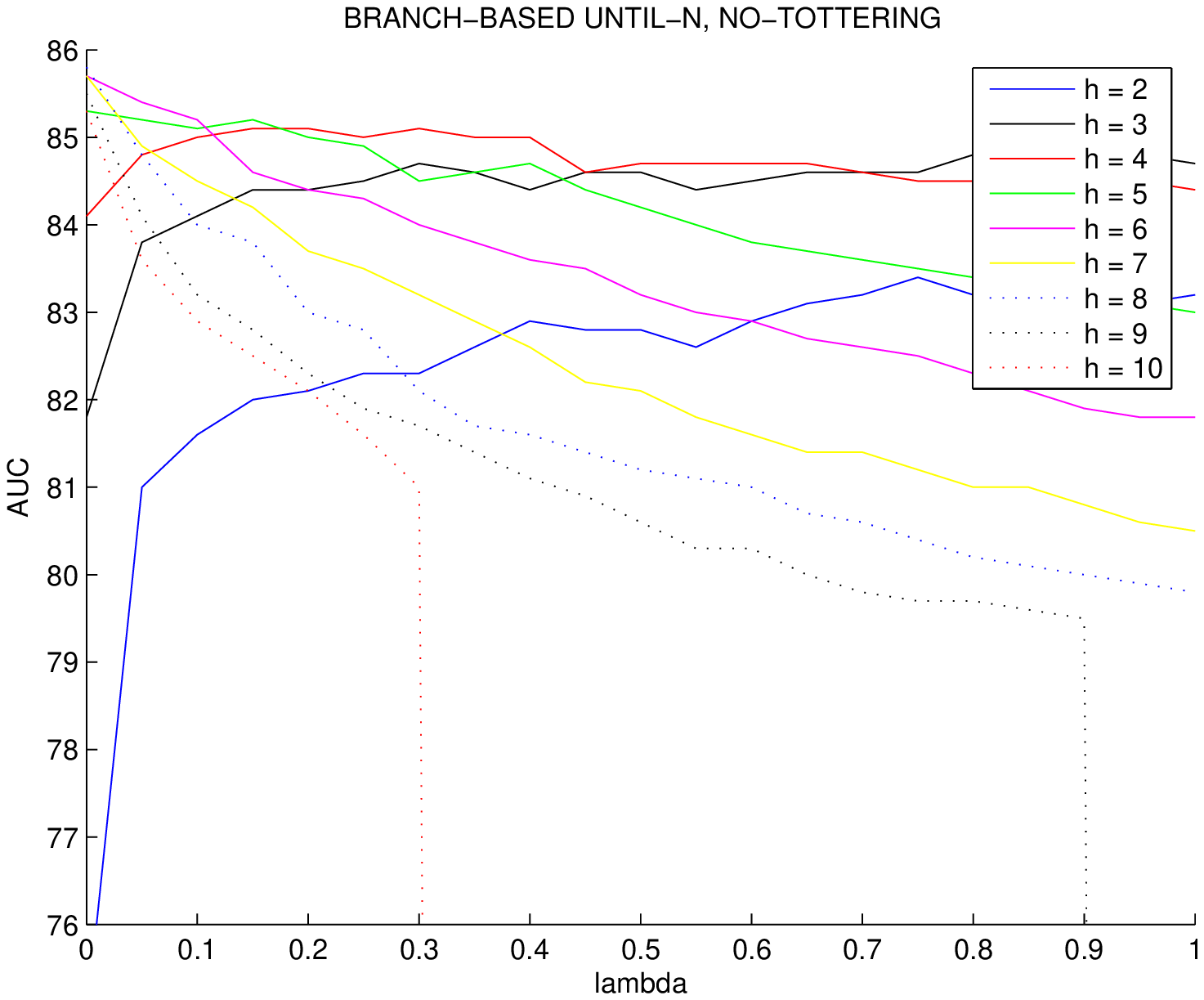}
\includegraphics[width=0.5\textwidth]{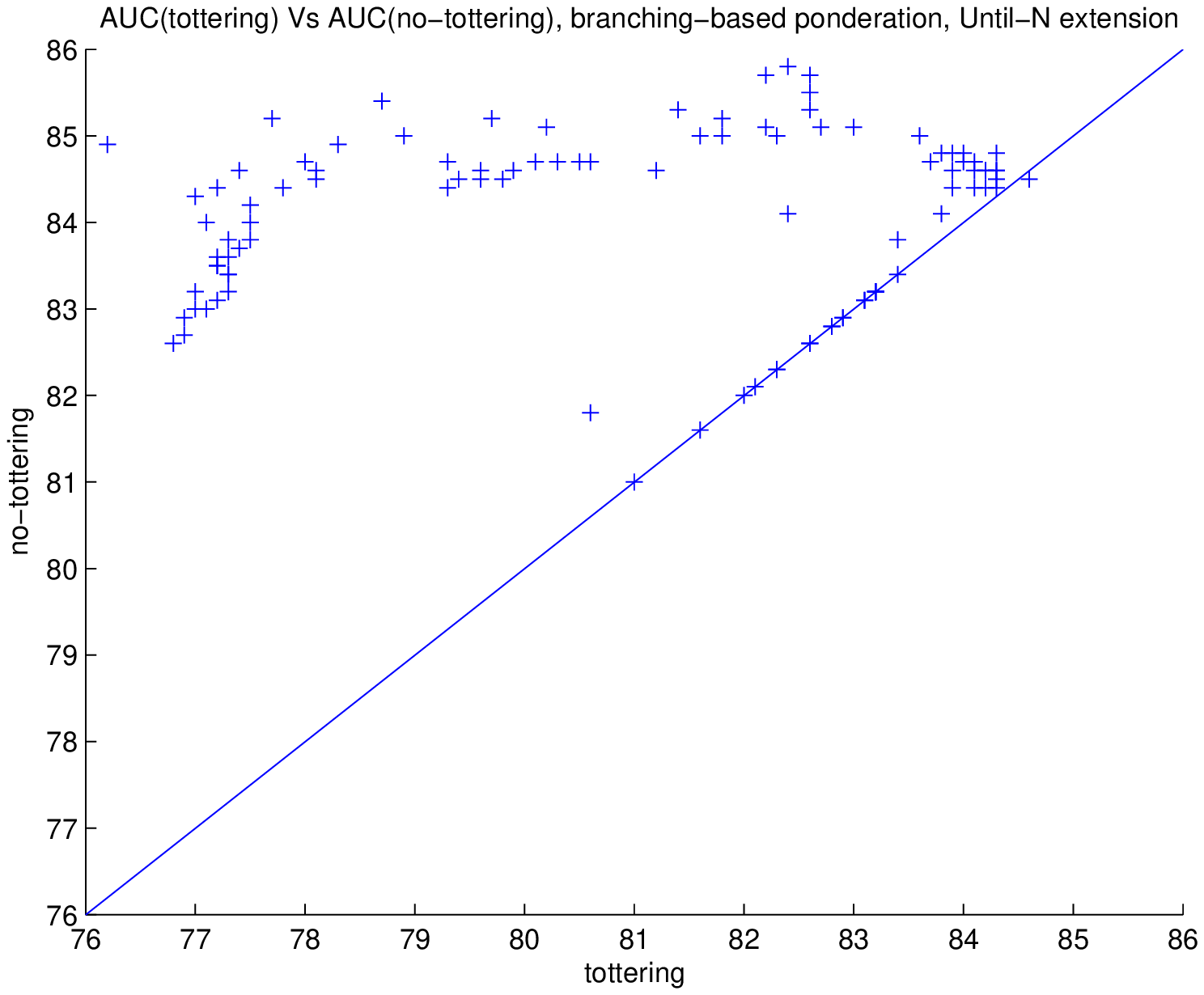} 
\caption{Second dataset. Left: evolution of the AUC with respect to $\lambda$ at different orders $h$ for the no-tottering extension (\ref{eq:kernel-nototter}) of the until-N branching-based kernel (\ref{eq:kernel-untilN}). Right: no-tottering AUC values Vs original values.}\label{fig:nototters-branchUntilN-mutag2}
\end{figure}

{\bf No-tottering extension: }\\
Figure \ref{fig:nototters-size-mutag2}, \ref{fig:nototters-branch-mutag2} and \ref{fig:nototters-branchUntilN-mutag2} respectively present the results of the no-tottering extension (\ref{eq:kernel-nototter}) of the size-based (\ref{eq:kernel-size}), branching-based (\ref{eq:kernel-branch}), and until-N branching-based (\ref{eq:kernel-untilN}) kernels.

Several observations are consistent with the first dataset.
We can likewise note that with the no-tottering extension, the introduction of tree-patterns is systematically beneficial in both kernels.
Moreover, at a given order, optimal results observed with the two kernels are similar, and the corresponding $\lambda$ value is smaller with the branching-based kernel.
Finally, the no-tottering extension limits the combinatorial explosion of the kernels computation.

There is however a striking difference because results are optimal here for patterns of order 3, patterns of order 2 rank second, and the results gradually decrease for orders greater than 3.
This behavior is exactly opposite to the one we observed with the first dataset, where results gradually increased with the order of the patterns and were optimal for patterns of order 8.
This therefore tends to confirm that distinct features are to be detected within the two datasets, and can be explained by the fact that the compounds are structurally similar in the first dataset, and different (or {\em noncongeneric}) in the second one.
Indeed, while the kernel needs to detect subtle differences between the compounds of the first dataset, it must identify regular patterns within the second one, and it is not surprising that discriminating patterns are shorter in this case.
This observation supports the intuition that the choice of the order of the patterns should to be related to (or learned from) the dataset itself, as suggested in section \ref{sec:gartner-def}.
Finally, we note that the best AUC value is around 84 \% (corresponding to a relative improvement of 7\% over the corresponding walk-based kernel), and is therefore similar to that obtained with the original formulation of the kernel.
Nevertheless, we observe from the curves on the right-hand side that contrary to the first dataset, the no-tottering extension has a limited overall impact.
This  is due to the surprising fact that here, the no-tottering extension does not seem to be beneficial by itself, since we can note that it systematically degrades the performance of the corresponding walk-based kernels, obtained for $\lambda = 0$.
As a result, even though  the introduction of tree-patterns is beneficial in both cases, better performances can be obtained here if we consider tottering tree-patterns.
Once again this behavior is opposite to that of the first dataset.
This might be explained as well by the fact that, contrary to the first dataset, the molecules considered here are structurally different, and as a result, tottering can help finding common features between these noncongeneric compounds.

Concerning the no-tottering extension  (\ref{eq:kernel-nototter}) of the until-N branching-based kernel  (\ref{eq:kernel-untilN}), results presented in Figure \ref{fig:nototters-branchUntilN-mutag2} are not clear.
Indeed, in that case, the introduction of the tree-patterns only improves the results for patterns of limited order, and for patterns of order greater than 4, results systematically decrease.
We can however note the interesting point that optimal results obtained for patterns of order 5 to 10 converge to a global optimal value between 85 and 86 \%.
This therefore tends to confirm that in the no-tottering case, the until-N extension can help solving the problem of pattern order selection by considering a  maximal pattern order large enough (here, $h >4$).
Nevertheless, the striking difference with the results obtained with the first dataset is that in this case, when $h > 4$,  the introduction of tree-patterns could not further improve the results obtained by the until-N walk-based kernel, that constitute the overall best performance we could observe for this dataset.

\section{Discussion}\label{sec:discussion}

This paper introduces a family of graph kernels based on the detection of common tree patterns in the graphs.
In a first step, we revisited an initial formulation presented in \cite{Ramon2003Expressivity}, from which we derived two kernels with explicit feature spaces and inner products.
A parameter $\lambda$ enters their definition and makes it possible to control the complexity of the features characterizing the graphs.
At the extreme, admissible tree-patterns consist of linear chains of graph vertices, and the kernels resume to a classical graph kernel based on the detection of common walks \citep{Gartner2003graph}.
Walk-based graph kernels are therefore generalized to a wider class of kernels defined by features of increasing levels of complexity.
In a second step we introduced two modular extensions to this initial formulation.
On the one hand, the set of trees initially indexing the feature space is enriched by the set of their subtrees with an {\em until-N} extension, leading to a wider and more general feature space.
On the other hand, a {\em no-tottering} extension prevents spurious tree-patterns to be detected, based on the notion of "tottering" initially introduced in the context of walk-based graph kernels \citep{Mahe2005Graph}. 

In the context of chemical applications, experiments on two toxicity datasets demonstrate that the tree-pattern graph kernels under their initial formulation improve over their walk-based counterpart.
However, while a significant improvement could be observed for relatively small patterns, experiments revealed the difficulty to handle high order patterns. 
This is due to the fact that the number of tree-patterns detected in the graphs increases exponentially with their depth, which leads to a combinatorial explosion of the kernels computation for large patterns.
For this reason, the until-N extension showed to be useless in this context: patterns of a given order are drowned within the flood of patterns of greater order, and the two kernel formulations turned out to be equivalent.
With the elimination of artificial tree-patterns, the no-tottering extension limits this combinatorial explosion, and patterns of higher order can be considered in the kernel.
This was in particular beneficial to the first dataset where optimal results were obtained with high-order no-tottering patterns.
Nevertheless, we notice that this extension is not always beneficial, and that in some cases, artificial common patterns due to the tottering phenomenon can help detecting molecular similarity.
This is in particular the case for the second dataset, and can be explained by the fact that, in opposition to the first dataset, it consists of structurally different compounds.
The combination of the two extensions led to mixed results. 
For the first dataset, we observe that the introduction of tree-patterns in this context could now improve over their walk-based counterparts for any maximum pattern order.
This suggests that the limitation of the combinatorial explosion offered by the no-tottering extension makes it possible to combine patterns of different order in the kernel.
However, albeit close, optimal results with the until-N extension could not come up with the optimal results that were obtained with no-tottering patterns of a given order. 
This suggests that very precise patterns were to be detected, and that their discriminative power is reduced by the addition of other, less predictive, patterns.
For the second dataset, the combination of the two extensions led to optimal results.
In that case however, the introduction of tree-patterns was not always beneficial and these optimal results were obtained by until-N, no-tottering walk-kernels.
Finally, we can note that, when the maximum order of the patterns considered is large enough,  results obtained with the until-N extension and no-tottering patterns tend to converge to a global optimum which is close, or equal to, to the overall best performance observed in both datasets.

Among the possible extensions to our work, we note that it might be relevant in the context of chemical applications to incorporate chemical knowledge in the graph representation of the molecules.
For instance, it is well known that physico-chemical properties of atoms are related to their position in the molecule, and as a first step in this direction, an enrichment of atom labels by their Morgan indices led to promising results in the context of walk-based kernels \citep{Mahe2005Graph}.
However, this particular approach is likely to have a lesser impact in this context, because the information encoded by the Morgan indices is at some extend already incorporated in the tree-patterns.
Alternatively, we note that the kernel implementation could easily be extended in order to introduce a flexible matching between tree-patterns based on measures of similarity between pairs of vertices and edges, following for instance the construction of the marginalized kernel between labeled graphs \citep{Kashima2004Kernels}.
Such an extension would induce an increase in the cost of computing the kernel, but is likely to make sense for chemical applications, where atoms of different types can exhibit similar properties.

\appendix

\section{Proof of Propositions \ref{prop:comput-size} and \ref{prop:comput-branch}}\label{app:proof1}
In Propositions \ref{prop:comput-size} and \ref{prop:comput-branch}, we want to prove that for the graphs $G_1$ and $G_2$
\begin{equation}\label{eq:proof1}
\sum_{t \in \mathcal{B}_h} w(t) \psi_t(G_1) \psi_t(G_2) =  \alpha(h) \sum_{u \in \mathcal{V}_{G_1}} \sum_{v \in \mathcal{V}_{G_2}} k_h(u,v),
\end{equation}
where in Proposition \ref{prop:comput-size}, $\alpha(h) = \lambda^{-h}$ and $w(t) = \lambda^{|t|-h}$, while in Proposition \ref{prop:comput-branch}, $\alpha(h) = 1$ and $w(t) = \lambda^{\text{branch}(t)}$.

From Definition \ref{def:tree-counting} we have $\displaystyle \psi_t(G) =  \sum_{u \in \mathcal{V}_G} \psi^{(u)}_t(G)$.
As a result, 
$$
\sum_{t \in \mathcal{B}_h} w(t) \psi_t(G_1) \psi_t(G_2) = \sum_{u \in \mathcal{V}_{G_1}} \sum_{v \in \mathcal{V}_{G_2}} \Big( \sum_{t \in \mathcal{B}_h} w(t) \psi^{(u)}_t(G_1) \psi^{(v)}_t(G_2)  \Big),
$$
and in order to prove (\ref{eq:proof1}) we just need to prove
\begin{equation}\label{eq:proof2}
\sum_{t \in \mathcal{B}_h} w(t) \psi^{(u)}_t(G_1) \psi^{(v)}_t(G_2) = \alpha(h)  k_h(u,v).
\end{equation}

        \subsection{Proof of Proposition \ref{prop:comput-size}}\label{sec:proof1}
In order to prove Proposition \ref{prop:comput-size}, it follows from (\ref{eq:proof2}) that we just need to prove that
\OMIT{\begin{align*}
\frac{1}{\lambda^h} k_h(u,v) & = \sum_{t \in \mathcal{B}_h} \lambda^{|t|-h} \psi^{(u)}_t(G_1) \psi^{(v)}_t(G_2) \\
                &= \frac{1}{\lambda^h}  \Big( \lambda {\bf 1}(l(u) = l(v)) \sum_{R \in \mathcal{M}(u,v)} \prod_{(u',v') \in R} k_{h-1}(u',v') \Big),
\end{align*}}
$$
\frac{1}{\lambda^h} k_h(u,v)  = \sum_{t \in \mathcal{B}_h} \lambda^{|t|-h} \psi^{(u)}_t(G_1) \psi^{(v)}_t(G_2)\,,
$$
or equivalently:
\OMIT{
\begin{align*}
k_h(u,v) & = \sum_{t \in \mathcal{B}_h} \lambda^{|t|} \psi^{(u)}_t(G_1) \psi^{(v)}_t(G_2) \\
        & = \lambda {\bf 1}(l(u) = l(v)) \sum_{R \in \mathcal{M}(u,v)} \prod_{(u',v') \in R} k_{h-1}(u',v'),
\end{align*}}
\begin{equation}\label{eq:tobeproved}
k_h(u,v)  = \sum_{t \in \mathcal{B}_h} \lambda^{|t|} \psi^{(u)}_t(G_1) \psi^{(v)}_t(G_2)\,,
\end{equation}
where $k_{h}$ is defined recursively by $k_{1}(u,v) = \lambda {\bf 1}(l(u) = l(v))$ and for $h>1$:
\begin{equation}\label{eq:kh}
k_{h}(u,v) = \lambda {\bf 1}(l(u) = l(v)) \sum_{R \in \mathcal{M}(u,v)} \prod_{(u',v') \in R} k_{h-1}(u',v') \,.
\end{equation}

We prove (\ref{eq:tobeproved}) by induction on $h$. The case $h=1$ is rather trivial. Indeed, a tree of depth one is just a single node, and $\psi^{(u)}_t(G_1)$ is therefore equal to $1$ if $l(u)=l(r(t))$, $0$ otherwise. It follows that 
\begin{align*}
\sum_{t \in \mathcal{B}_1} \lambda^{|t|} \psi^{(u)}_t(G_1) \psi^{(v)}_t(G_2) & = \sum_{t \in \mathcal{B}_1} \lambda {\bf 1}( l(r(t)) = l(u)) {\bf 1}( l(r(t)) = l(v)) \\                                                                        & = \lambda {\bf 1}( l(u) = l(v)),
\end{align*}
which corresponds to $k_1(u,v)$.

Let us now assume that (\ref{eq:tobeproved}) is true at order $h-1$, and let us prove that it is then also true at order $h>1$. Combining the recursive definition of $k_{h}$ (\ref{eq:kh}) with the induction hypothesis (\ref{eq:tobeproved}) at level $h-1$ we first obtain:
\begin{equation}\label{eq:jp1}
k_{h}(u,v) = \lambda {\bf 1}(l(u) = l(v)) \sum_{R \in \mathcal{M}(u,v)} \prod_{(u',v') \in R} \sum_{t' \in \mathcal{B}_{h-1}} \lambda^{|t'|} \psi^{(u')}_{t'}(G_1) \psi^{(v')}_{t'}(G_2)\,.
\end{equation}
Second, for any graph $G$, let us denote by $\mathcal{P}_n^{(u)}(G)$ the set of balanced tree-patterns of order $n$ rooted in $u \in \mathcal{V}_G$, and for any tree-pattern $p\in\mathcal{P}_n^{(u)}(G)$ let $t(p) \in\mathcal{B}_{n}$ denote the corresponding tree. With these notations we can rewrite, for any $n\geq 1$ and $(u,v)\in G_{1}\times G_{2}$:
\begin{equation}\label{eq:jp2}
\sum_{t \in \mathcal{B}_n} \lambda^{|t|} \psi^{(u)}_t(G_1) \psi^{(v)}_t(G_2) = \sum_{p_{1} \in \mathcal{P}_{n}^{(u)}(G_1)} \sum_{p_{2} \in \mathcal{P}_{n}^{(v)}(G_2)} \lambda^{|t(p_{1})|}{\bf 1}(t(p_{1}) = t(p_{2})).
\end{equation}
Indeed both sides of this equation count the number of pairs of similar tree-patterns rooted in $u$ and $v$. Plugging (\ref{eq:jp2}) into (\ref{eq:jp1}) we get:
\begin{equation}\label{eq:jp3}
k_{h}(u,v) = \lambda {\bf 1}(l(u) = l(v)) \sum_{R \in \mathcal{M}(u,v)} \prod_{(u',v') \in R} \sum_{p_{1} \in \mathcal{P}_{h-1}^{(u')}(G_1)} \sum_{p_{2} \in \mathcal{P}_{h-1}^{(v')}(G_2)} \lambda^{|t(p_{1})|}{\bf 1}(t(p_{1}) = t(p_{2}))\,.
\end{equation}
Now we use the fact that any tree-pattern $p$ of order $h$ can be uniquely decomposed into a tree-pattern $p'$ of order $2$ and a set of tree-patterns of order $h-1$ rooted at the leaves of $p'$. 
We note that matching two tree-patterns is equivalent to matching the tree-patterns in their decomposition, and that the sets of leaves of tree-patterns of order 2 rooted respectively in $u$ and $v$ matching each other are exactly given by $\Mcal(u,v)$. 
In other words, (\ref{eq:jp3}) performs a summation over pairs of matching tree-patterns of depth $h$, rooted respectively in $u$ and $v$: the corresponding pairs of patterns of order 2 are implicitly matched by the summation over  $\Mcal(u,v)$ and the condition ${\bf 1}(l(u) = l(v))$, and the subsequent pairs of patterns $(p_1,p_2)$ of order $h-1$ are matched by the product of conditions ${\bf 1}(t(p_{1}) = t(p_{2}))$.

The tree-pattern $p_1$  in $G_{1}$ of such a matching pair of tree-patterns of order $h$ rooted in $(u,v)$ decomposes as a pattern of depth 2 rooted in $u$ with leaves in some $R \in \mathcal{M}(u,v)$, and a set of patterns $p_{1}(u')$ of depth $h-1$ rooted in the leaves $u'\in R$. 
By (\ref{eq:jp3}), to each such matching pair is associated the weight $\lambda\times\prod_{(u',v')\in R} \lambda^{|t(p1(u'))|}$, which is exactly equal to $\lambda^{|t(p_1)|}$ since we obviously have $|t(p_1)| = 1 +\sum_{(u',v') \in R} |t(p_{1}(u'))| $.
As a result, (\ref{eq:jp3}) can be rewritten as:
$$
k_{h}(u,v) = \sum_{p_{1} \in \mathcal{P}_{h}^{(u)}(G_1)} \sum_{p_{2} \in \mathcal{P}_{h}^{(v)}(G_2)} \lambda^{|t(p_{1})|}{\bf 1}(t(p_{1}) = t(p_{2})),
$$
which combined with (\ref{eq:jp2}) proves (\ref{eq:tobeproved}).
\qed

        \subsection{Proof of Proposition \ref{prop:comput-branch}}
The proof of Proposition \ref{prop:comput-branch} is a straightforward variant of the proof of Proposition \ref{prop:comput-size}. By (\ref{eq:proof2}) we need to show that
\begin{equation}\label{eq:tobeproved10}
k_h(u,v)  = \sum_{t \in \mathcal{B}_h} \lambda^{\text{branch}(t)} \psi^{(u)}_t(G_1) \psi^{(v)}_t(G_2)\,,
\end{equation}
where $k_{h}$ is defined recursively by $k_{1}(u,v) = {\bf 1}(l(u) = l(v))$ and for $h>1$:
\begin{equation}\label{eq:kh10}
k_{h}(u,v) = \frac{{\bf 1}(l(u) = l(v))}{\lambda} \sum_{R \in \mathcal{M}(u,v)} \prod_{(u',v') \in R} \lambda k_{h-1}(u',v') \,.
\end{equation}
We proceed again by induction over $h$ to prove (\ref{eq:tobeproved10}). 
The case $h=1$ is easily done by checking, using an argument similar to that of the previous proof, that (\ref{eq:tobeproved10}) is one if $l(u)$ and $l(v)$ are identical, zero otherwise, which corresponds to the definition of $k_1(u,v)$.
 If we assume that (\ref{eq:tobeproved10}) is true at the level $h-1$, we can plug it in (\ref{eq:kh10}) to obtain:
\begin{equation}\label{eq:jp1-bis}
k_{h}(u,v) = \frac{{\bf 1}(l(u) = l(v))}{\lambda} \sum_{R \in \mathcal{M}(u,v)} \prod_{(u',v') \in R} \sum_{t' \in \mathcal{B}_{h-1}} \lambda^{1+\text{branch}(t')} \psi^{(u')}_{t'}(G_1) \psi^{(v')}_{t'}(G_2)\;.
\end{equation}


We can then follow exactly the same line of proof as in the previous section and obtain the following equations 
\begin{equation}\label{eq:jp2-bis}
\sum_{t \in \mathcal{B}_n} \lambda^{\text{branch}(t)} \psi^{(u)}_t(G_1) \psi^{(v)}_t(G_2) = \sum_{p_{1} \in \mathcal{P}_{n}^{(u)}(G_1)} \sum_{p_{2} \in \mathcal{P}_{n}^{(v)}(G_2)} \lambda^{\text{branch}(t(p_{1}))}{\bf 1}(t(p_{1}) = t(p_{2})) \,,
\end{equation}
and
\begin{equation}\label{eq:jp3-bis}
k_{h}(u,v) = \frac{{\bf 1}(l(u) = l(v))}{\lambda} \sum_{R \in \mathcal{M}(u,v)} \prod_{(u',v') \in R} \sum_{p_{1} \in \mathcal{P}_{h-1}^{(u')}(G_1)} \sum_{p_{2} \in \mathcal{P}_{h-1}^{(v')}(G_2)} \lambda^{1 + \text{branch}(t(p_1))}{\bf 1}(t(p_{1}) = t(p_{2}))\,,
\end{equation}
that correspond respectively to (\ref{eq:jp2}) and (\ref{eq:jp3}).
The only difference with the previous proof is in the exponent of $\lambda$ to form the weight of a matching pair of tree-patterns.
By analogy with the previous proof, we consider the tree-pattern $p_1$ in $G_{1}$ of a pair of matching tree-patterns of depth $h$ rooted in $(u,v)$, that decomposes as a pattern of depth 2 rooted in $u$ with leaves in some $R \in  \mathcal{M}(u,v)$, and a set of patterns $p_{1}(u')$ of depth $h-1$ rooted in the leaves $u'\in R$. 
By (\ref{eq:jp3-bis}), to each such matching pair is associated the weight $\frac{1}{\lambda} \prod_{(u',v')\in R} \lambda^{1+\text{branch}(t(p_1(u')))} =   \lambda^{ -1 +  \sum_{(u',v')\in R} 1+\text{branch}(t(p_1(u'))) }$.
We observe that the number of leaves of a tree $t$, that we note $\text{leaves}(t)$, is equal to $1+\text{branch}(t)$.
The weight associated to the above pair of matching tree-patterns can therefore be written as  $ \lambda^{ -1 +  \sum_{(u',v')\in R} \text{leaves}(t(p_1(u')))}$.
Finally, because the number of leaves of the tree-pattern $p_1$ is equal to the sum of the leaves of the patterns $p_1(u')$,  it follows that this expression is equal to $\lambda^{-1 + \text{leaves}(t(p_1))} = \lambda^{\text{branch}(t(p_1))}$.
 As a result, we can write (\ref{eq:jp3-bis}) as
$$
k_{h}(u,v) = \sum_{p_{1} \in \mathcal{P}_{h}^{(u)}(G_1)} \sum_{p_{2} \in \mathcal{P}_{h}^{(v)}(G_2)} \lambda^{\text{branch}(t(p_1))}{\bf 1}(t(p_{1}) = t(p_{2})),
$$
which, combined with (\ref{eq:jp2-bis}), concludes the proof.
\qed

\section{Proof of Proposition \ref{prop:untilN}}\label{app:proof-untilN}

The proof presented in this section is very similar to the proofs of Propositions \ref{prop:comput-size} and \ref{prop:comput-branch}.
Based on the observations made in the beginning of Appendix \ref{app:proof1}, it follows from (\ref{eq:proof2}) that in order to prove Proposition \ref{prop:untilN}, we just need to prove that
\begin{equation}\label{eq:tobeproved11}
k_h(u,v) = \sum_{t \in \mathcal{T}_h} \lambda^{\text{branch}(t)} \psi^{(u)}_t(G_1) \psi^{(v)}_t(G_2) \, ,
\end{equation}
where $k_h$ is defined recursively by $k_{1}(u,v) = {\bf 1}(l(u) = l(v))$ and for $h>1$
\begin{equation}\label{eq:kh11}
k_h(u,v) = {\bf 1}(l(u) = l(v)) \Big( 1 + \sum_{R \in \mathcal{M}(u,v)} \frac{1}{\lambda} \prod_{(u',v') \in R} \lambda k_{h-1}(u',v') \Big).
\end{equation}
We proceed again by induction over $h$ to prove (\ref{eq:tobeproved11}). 
The case $h=1$ directly follows from the proof of Proposition \ref{prop:comput-branch}.
 If we assume that (\ref{eq:tobeproved11}) is true at the level $h-1$, we can plug it in (\ref{eq:kh11}) to obtain:
\begin{equation}\label{eq:jp1-ter}
k_h(u,v) = {\bf 1}(l(u) = l(v)) \Big( 1 + \sum_{R \in \mathcal{M}(u,v)} \frac{1}{\lambda} \prod_{(u',v') \in R} \sum_{t' \in \mathcal{T}_{h-1}} \lambda^{1+\text{branch}(t')} \psi^{(u')}_{t'}(G_1) \psi^{(v')}_{t'}(G_2) \Big).
\end{equation}
By analogy with the construction of the previous proof, for any graph $G$, let us denote by $\mathcal{P}_n^{(u)}(G)$ the set of  tree-patterns of depth 1 to $n$ rooted in $u \in \mathcal{V}_G$, and for any tree-pattern $p\in\mathcal{P}_n^{(u)}(G)$ let $t(p) \in\mathcal{T}_{n}$ denote the corresponding tree. 
Note that $\mathcal{P}_n^{(u)}(G)$ corresponds here to  general tree-patterns of depth 1 to $n$, in opposition to the balanced-tree patterns of order $n$ involved in the previous proofs.
With these notations we obtain similarly, for any $n\geq 1$ and $(u,v)\in G_{1}\times G_{2}$:
\begin{equation}\label{eq:jp2-ter}
\sum_{t \in \mathcal{T}_n} \lambda^{\text{branch}(t)} \psi^{(u)}_t(G_1) \psi^{(v)}_t(G_2) = \sum_{p_{1} \in \mathcal{P}_{n}^{(u)}(G_1)} \sum_{p_{2} \in \mathcal{P}_{n}^{(v)}(G_2)} \lambda^{\text{branch}(t(p_{1}))}{\bf 1}(t(p_{1}) = t(p_{2})),
\end{equation}
and, plugging (\ref{eq:jp2-ter}) into (\ref{eq:jp1-ter}), we get:
\begin{equation}\label{eq:jp3-ter}
\begin{split}
 k_h(u,v) =  & {\bf 1}(l(u) = l(v))  \\ 
& \times \Big( 1 + \sum_{R \in \mathcal{M}(u,v)} \frac{1}{\lambda} \prod_{(u',v') \in R} \sum_{p_{1} \in \mathcal{P}_{h-1}^{(u')}(G_1)} \sum_{p_{2} \in \mathcal{P}_{h-1}^{(v')}(G_2)} \lambda^{1+\text{branch}(t(p_{1}))}{\bf 1}(t(p_{1}) = t(p_{2})) \Big),
\end{split}
\end{equation}
which can be further decomposed into:
\begin{equation}\label{eq:jp4-ter}
\begin{split}
k_h(u,v) = &  {\bf 1}(l(u) = l(v)) \\
&  + \frac{{\bf 1}(l(u) = l(v)) }{\lambda}  \sum_{R \in \mathcal{M}(u,v)}  \prod_{(u',v') \in R} \sum_{p_{1} \in \mathcal{P}_{h-1}^{(u')}(G_1)} \sum_{p_{2} \in \mathcal{P}_{h-1}^{(v')}(G_2)} \lambda^{1+\text{branch}(t(p_{1}))}{\bf 1}(t(p_{1}) = t(p_{2})) .
\end{split} 
\end{equation}
The second part of the right member of (\ref{eq:jp4-ter}) matches pairs of tree-patterns of depth 2 to $n$ rooted in $(u,v)$.
It follows directly from the proof of Proposition \ref{prop:comput-branch} that such a pair $(p_1,p_2)$ of matching tree-patterns   is weighted by $\lambda^{\text{branch}(t(p_1))}$.
The first part of the right member of (\ref{eq:jp4-ter}) matches the trivial pair of tree-patterns of depth 1 rooted in $(u,v)$ consisting of the single nodes $(u,v)$.
The corresponding tree has a zero branching cardinality, and we can therefore write
$$
{\bf 1}(l(u) = l(v)) = \sum_{t \in \mathcal{T}_1} \lambda^{\text{branch}(t)} \psi_t^{(u)}(G_1) \psi_t^{(v)}(G_2). 
$$
Taken together, these two arguments show that (\ref{eq:jp4-ter}) can be written as
$$
k_h(u,v) = \sum_{t \in \mathcal{T}_h} \lambda^{\text{branch}(t)} \psi_t^{(u)}(G_1) \psi_t^{(v)}(G_2),
$$
which concludes the proof.
\qed

\section{Proof of Proposition \ref{prop:nototter-kernel}}\label{app:proof-nototter}

The proof is derived from results presented in \cite{Mahe2005Graph}.
The sets of walks and no-tottering walks of the graph $G = (\mathcal{V}_G, \mathcal{E}_G)$ are respectively defined by $\mathcal{W}(G) = \bigcup_{n=0}^{\infty} \mathcal{W}_n(G)$ and $\mathcal{W}^{NT}(G) = \bigcup_{n=0}^{\infty} \mathcal{W}_n^{NT}(G)$,  where
\begin{equation*}
\mathcal{W}_n(G) = \{ (v_0,\dots,v_n) \in \mathcal{V}_G^{n+1}: (v_i,v_{i+1}) \in \mathcal{E}_G, \; 0 \leq i \leq n-1 \}
\end{equation*}
is the set of walks of length $n$ defined is Section \ref{sec:gartner-def}, and 
\begin{equation*}
\mathcal{W}_n^{NT}(G) = \{ (v_0,\dots,v_n) \in \mathcal{W}_n(G) : \; v_i \neq v_{i+2}, \; 0 \leq i \leq n-2 \}
\end{equation*}
is the set of no-tottering walks of length $n$ defined in \cite{Mahe2005Graph}.
We start by stating the following lemma.
\begin{lemma}\label{lemma:nototters}
A tree-pattern $p$ of the graph $G$ associated to the tree $t$ is no tottering if, and only if, any walk  of G defined as a succession of vertices of $p$ corresponding to nodes of $t$ forming a path from its root to one of its leaves is no-tottering.
\end{lemma}

\begin{proof}[Proof of Lemma \ref{lemma:nototters}]
According to Definition \ref{def:nototter-counting}, let $(v_1,\dots,v_{|t|}) \in \mathcal{V}_G^{|t|}$ be a no-tottering tree pattern of the graph $G = (\mathcal{V}_G,\mathcal{E}_G)$ corresponding to the tree $t = (\mathcal{V}_t,\mathcal{E}_t)$, where  $\mathcal{V}_t = (n_1,\dots ,n_{|t|})$.
Let $(n_{i_0},\dots ,n_{i_k}) \in \mathcal{V}_t^{k+1}$ be a path from the root of $t$ to one of its leaves.
By Definition  \ref{def:tree-pattern}, it is clear that $(v_{i_0},\dots ,v_{i_k}) \in \mathcal{W}(G)$.
Moreover, by the definition of paths  we have $(n_{i_m}, n_{i_{m+1}}), (n_{i_{m+1}},n_{i_{m+2}}) \in \mathcal{E}_t$ for $0\leq m \leq k-2$.
By Definition \ref{def:nototter-counting}, this implies that $v_{i_{m}} \neq v_{i_{m+2}}$ for $0 \leq m \leq k-2$, meaning that $(v_{i_0},\dots,v_{i_k}) \in \mathcal{W}^{NT}(G)$.
Conversely, let $ p \in \mathcal{V}_G^{|t|}$ be a tree-pattern of the graph  $G = (\mathcal{V}_G,\mathcal{E}_G)$ corresponding to the tree $t = (\mathcal{V}_t,\mathcal{E}_t)$.
Consider the set of walks of $G$ defined as successions of vertices of $p$ associated to nodes of $t$ forming paths from its root to its leaves.
If these walks are not tottering, it is clear from Definition \ref{def:nototter-counting} that the tree-pattern itself is not tottering.
\end{proof}

We can now state the proof of Proposition \ref{prop:nototter-kernel}.
\begin{proof}[Proof of Proposition \ref{prop:nototter-kernel}]
If, according to Definition \ref{def:transfo}, we let $G'$ be the transformed graph of $G$,  \cite{Mahe2005Graph} showed that there is a bijection between $\mathcal{W}^{NT}(G)$ and the set of walks of $G'$ starting in a vertex corresponding to a vertex of $G$, which can be formally defined as
$$\mathcal{W}^{\{V_G\}}(G') = \{ (v_0,\dots,v_n) \in \mathcal{W}(G') : \; v_0 \in \{V_G\},  n \in \mathbb{N} \},$$
if we let $V_G \subset \mathcal{V}_{G'}$ be the subset of $\mathcal{V}_{G'}$ that corresponds to $\mathcal{V}_G$.
It follows from Lemma \ref{lemma:nototters} that there is a bijection between the set of no-tottering tree-patterns of $G$  and the set of tree-patterns of $G'$ rooted in a vertex of $V_G$.
Finally, \cite{Mahe2005Graph} showed that a walk in $\mathcal{W}^{NT}(G)$ and its image in $\mathcal{W}^{\{V_G\}}(G')$ are identically labeled, which enables to count no-tottering labeled walks in $G$, by counting identically labeled walks in $G'$ starting in a vertex of $V_G$ .
It follows that counting no-tottering tree-patterns in $G$ is equivalent to counting tree-patterns in $G'$ rooted in a vertex of $V_G$.
As a result, we have $\psi_t^{NT}(G) = \psi_t^{\{V_G\}}(G')$, which concludes the proof.
\end{proof}


\end{document}